\documentclass[useAMS,usenatbib,usegraphicx]{mn2e}

\title[Ultraviolet dust attenuation in galaxies]
{Effects of dust scattering albedo and 2175 \AA\ bump on ultraviolet
colours of normal disc galaxies}
\author[A. K. Inoue et al.]{Akio K. Inoue$^{1,2}$\thanks{E-mail: 
akinoue@las.osaka-sandai.ac.jp (AKI).}, 
Veronique Buat$^{2}$, Denis Burgarella$^{2}$, Pasquale Panuzzo$^{3}$, 
\newauthor 
Tsutomu T. Takeuchi$^{2}$\thanks{Present address: Astronomical
Institute, Tohoku University, Aoba, Aramaki, Aoba-ku, Sendai 980--8578,
JAPAN} 
and Jorge Iglesias-P{\'a}ramo$^{4}$\\
$^{1}$College of General Education, Osaka Sangyo University, 
3-1-1, Nakagaito, Daito, Osaka 574-8530, Japan\\
$^{2}$Laboratoire d'Astrophysique de Marseille, 
Traverse du Siphon, BP 8, 13376 Marseille, CEDEX 12, France\\
$^{3}$INAF Padova, Vicolo dell'Osservatorio 5, I-35122 Padova, Italy\\
$^{4}$Instituto de Astrof{\'i}sica de Andaluc{\'i}a (CSIC), 
18008 Granada, Spain}
\begin{document}

\date{Accepted. Received; in original form}

\pagerange{\pageref{firstpage}--\pageref{lastpage}} \pubyear{2005}

\maketitle

\label{firstpage}

\begin{abstract}
 We discuss dust properties in the interstellar medium (ISM) of
 nearby normal galaxies, by comparing observations in the ultraviolet
 (UV) with simulations by a radiative transfer model. 
 The observed UV colours of nearby galaxies show a
 reddening relative to their expected intrinsic colours. Some authors
 argued that the Milky Way dust cannot reproduce the reddening because
 of the prominent 2175 \AA\ absorption bump. Other authors proposed a
 reduction mechanism of the bump strength in an {\it attenuation law}
 derived from the ratio of the observed intensity to the intrinsic one
 through an age-selective attenuation (i.e., young stars are more
 attenuated selectively). We newly find that the wavelength dependence
 of the scattering albedo also has a strong effect on the UV colour; an
 albedo decreasing toward shorter wavelengths (except for the
 absorption bump range) produces a significant UV reddening. 
 After comparing the observed UV colours of nearby normal galaxies with
 those expected from radiative transfer simulations assumed several dust
 models, we find two sorts of dust suitable for these galaxies: 
 (1) dust with a bump and a smaller albedo for a shorter wavelength
 (except for the bump range), and (2) dust without any bump but with an 
 almost constant albedo. If very small carbonaceous grains responsible
 for the common unidentified infrared emission band are also the bump
 carrier, the former dust is favorable. Finally, we derive mean
 attenuation laws of various dust models as a function of the UV
 attenuation, and derive some relations between the UV attenuation and
 observable/theoretical quantities.
\end{abstract}


\section{Introduction}

Dust grains strongly affect our observations, especially in the
ultra-violet (UV) band, through {\it extinction} (absorption and
scattering processes). When we observe galaxies whose stars are not
resolved by a telescope, the effects of the radiative transfer, i.e., 
multiple scatterings and the configuration of dust grains and stars, 
are also effective. In this paper, we call the effective extinction
including such effects {\it attenuation}\footnote{Other words are also
used in the literature: obscuration, effective extinction, and so
on. See discussions in \cite{cal01}.}.
To correctly understand the intrinsic spectrum of galaxies (unresolved
into individual stars), we must correct observational data for the
dust {\it attenuation} not for the {\it extinction}. 

Theoretically, to know the dust attenuation through a galactic disc is
to solve the equation of the radiative transfer through the disc, 
assuming the wavelength dependence of average dust properties of
absorption and scattering, i.e., the {\it extinction law}. 
The most prominent feature in the average extinction law of the Milky
Way (MW) is the absorption ``bump'' at 2175 \AA. We find this feature
toward almost all lines of sight in the interstellar medium (ISM) of the
MW \citep[e.g.,][]{fit99}, suggesting that the carrier of the bump is
quite common in the ISM of the MW. We also find the bump in the average
extinction law of the Large Magellanic Cloud (LMC) but not in the
average extinction law of the Small Magellanic Cloud (SMC) 
\citep[e.g.,][]{whi03}.

Observationally, the wavelength dependence of the attenuation amount,
i.e., the {\it attenuation law}, has been obtained for nearby UV bright
starburst galaxies observed with the {\it IUE} satellite \citep{cal94}.
In an average attenuation law of these galaxies (so-called the Calzetti
law), the bump is very weak or absent \citep{cal94}, although some
galaxies in the sample show a sign of the bump in their spectra
\citep{nol05}. Based on a radiative transfer model, \cite{gor97} and
\cite{wg00} argued that this lack of the bump in the Calzetti law is
a sign of the absence of the bump in the extinction law, suggesting 
the absence of the bump carrier in the starburst region. On the other
hand, with another radiative transfer model, GRASIL \citep{sil98},
\cite{gra00} pointed out that, even if the bump exists in the extinction
law, the strength of the bump can be greatly reduced in the attenuation
law by a radiative transfer effect coupled with an 
{\it age-selective attenuation}, i.e. young stars are more attenuated
selectively.

The {\it IUE} starburst galaxies also show a tight correlation between
the observed UV spectral slope ($\beta$; 
$f_\lambda \propto \lambda^{-\beta}$) and the infrared (IR)-to-UV flux
ratio (so-called IRX): a redder $\beta$ for a larger IRX
\citep{cal94,meu99}. 
Since the IRX well relates to the UV attenuation \citep{bua96}, the 
correlation means that the UV spectrum (i.e. the UV colour) becomes
redder monotonically as the attenuation increases. \cite{wg00} argued
that the UV colour cannot be redden by the extinction law of the MW 
even if the dust column density of a medium increases. 
This is because the absorption bump lies in the near-UV (NUV). 
Indeed, the extinction in the NUV is slightly larger than that in
the far-UV (FUV) for the MW extinction law. Their finding again suggests
the lack of the bump carrier in the starburst galaxies.

A suggested candidate of the bump carrier is very small
carbonaceous grains like PAHs (Polycyclic Aromatic Hydrocarbons;
\citealt{leg89}), QCCs (Quenched Carbonaceous Composites;
\citealt{sak83,wad99}), and UV processed HACs (Hydrogenerated Amorphous
Carbon grains; \citealt{men98}), 
although this is not settled yet \citep{whi03,dra03a,hen04}. 
These very small carbonaceous particles are confidently attributed to
the unidentified infrared (UIR) emission band in 3--13 \micron\ 
\citep{leg84,sak84,whi03}. The UIR emission band is quite common in the
ISM of the MW \citep[e.g.,][]{ona04} and of other galaxies
\citep[e.g.,][]{gen00}, except for low-metallicity ($\la 1/5$
$Z/Z_\odot$) galaxies in which the UIR emission is weak or absent 
\citep{eng05}. If the very small carbonaceous grains producing the
UIR emission are really responsible for the bump, we should find the
bump in the extinction law of other galaxies (but not so low-metallicity). 
Indeed, the bump found in M31 \citep{bia96} and some distant galaxies,
for example, a lensing galaxy at $z=0.83$ \citep{mot02} and Mg {\sc ii}
absorption systems at $z=1.5$ \citep{wan04}. There are also signs of the
bump imprinted in the observed UV spectra of a galaxy at $z=0.048$
\citep{bur05a}, of some {\it IUE} starburst galaxies \citep{nol05}, 
and of some star-forming galaxies at $z\sim2$ \citep{nol05}.

The amount of the bump carrier may depend on the star-forming activity 
\citep[ and references therein]{gor05}. 
\cite{gor98} found one sight line with the bump in the
SMC. Interestingly, this sight line is toward a quiescent area in the
SMC. \cite{val03} found one sight line without the bump toward an
actively star-forming region (Trumpler 37) in the MW 
\citep[see also][]{sof05}. \cite{whi04} also 
found a sight line without the bump toward a molecular cloud in the MW
\citep[see also][]{sof05}. Furthermore, the UIR emission flux (possibly
related to the bump carrier) relative to the far-IR flux decreases as
the intensity of the interstellar radiation field increases
\citep{ona04}. Such a bump carrier fragile against the star-forming
activity may explain the absence of the bump in the Calzetti law.
On the other hand, there is a good correlation between the UIR emission
strength and the star-forming activity for starburst galaxies as well as
for normal galaxies \citep{gen00}.

Quiescent or modest star-forming ``normal'' galaxies and 
ultra-luminous infrared galaxies (ULIRGs) do not follow the tight
correlation of the {\it IUE} starburst galaxies on the IRX-$\beta$
diagram \citep{bel02,gol02}. Particularly, normal galaxies show 
systematically redder UV colours than those of the {\it IUE} starburst
galaxies \citep{bel02,kon04}. This fact has recently been confirmed by
the {\it GALEX} satellite \citep{mar05} for larger samples of nearby
galaxies selected in the NUV or optical \citep{bua05,sei05}. 
Resolved star-forming regions (aperture size of 520 pc) in M51 also show
the same trend as the normal galaxies on the IRX-UV colour diagram 
\citep{cal05}.
According to \cite{wg00}, the MW type dust can not reproduce even the UV
colour of the {\it IUE} starburst galaxies, much less the redder UV
colour of normal galaxies. Does the red UV colour of normal galaxies
indicate the lack of the bump carrier in their ISM and suggest different
origins of the bump and the UIR emission?

With a simple power-law type attenuation law (i.e. without a bump) 
as introduced by \cite{cha00}, \cite{kon04} claimed that the redder UV
colour of normal galaxies is due to a post-burst stellar population if
galaxies have an intermittent star formation. They expected a trend
that the current star formation rate relative to the past average one
(so-called the birth-rate parameter) depends on the distance from the
starburst relation on the IRX--$\beta$ diagram. However, the observed
trend is weak \citep{cor05,pan05}. In addition, the expected recent
burst of nearby galaxies observed with the {\it GALEX} seems to be too
weak to change their UV colours significantly \citep{bur05b}. 

\cite{bur05b} applied a more realistic attenuation law, i.e., a
power-law plus a Gaussian bump to their statistical investigation in
order to understand the nature of {\it GALEX} galaxies. They found
that an attenuation law with a bump and a somewhat steep slope is
suitable for these galaxies. Such a steep attenuation law is expected
from an {\it age-selective attenuation} \citep{ino05}. More recently, an
updated GRASIL model \citep{pan05} very well reproduced the red UV
colours of {\it GALEX} galaxies with the MW type dust. They adopted
a more realistic stellar distribution; younger stars are more deeply
embedded in the dust disc, whereas older stars distribute more
extensively \citep[e.g.,][]{rob03,zar04}. This realistic configuration
of dust and stars depending on the stellar age produces an 
{\it age-selective attenuation}. This results in a steep attenuation law
which overcomes the {\it blueing} by the bump.

In addition to the {\it age-selective attenuation}, this paper newly
discusses the effect of the wavelength dependence of the scattering
albedo on the UV colour. In fact, the dust properties adopted by
\cite{wg00} and the GRASIL are different from each other, especially the
wavelength dependence of albedos. This point significantly affects 
the expected UV colour as shown later (\S3). \cite{wg00} empirically
derived the wavelength dependence of the albedo from a large compilation
of the albedos estimated from observations with radiative transfer
models \citep[ for a review]{gor04}. On the other hand, the dust
properties adopted in the GRASIL are a theoretical model by Draine and
co-workers \citep{dra03,wei01}. Such a difference in the adopted dust
models could play a role in the contradictory conclusions from the two
groups.

This paper thoroughly examines dust properties in nearby galaxies, 
in particular the presence of the bump and the wavelength dependence
of the albedo, based on the {\it GALEX} colour. We adopt a
one-dimensional plane-parallel radiative transfer model developed by
\cite{ino05}. While its computational geometry is one-dimensional, this
model can treat the clumpiness of stars and dust thanks to the
mega-grain approximation \citep{var99}. The validity of this
approximation has been clearly shown by \cite{var99} who compared the
approximation with a three-dimensional Monte Carlo radiative transfer.
Owing to the computational cheapness of the one-dimensional
calculation, we can investigate a very wide range of physical quantities
of disc galaxies.

In the next section, we present a description of the radiative transfer
model, detailed explanations of the dust models, and the set up of the
plane-parallel discs. The main results and discussions are presented in
section 3 where we search suitable dust models for nearby
``normal'' galaxies. In section 4, we derive mean attenuation laws as a
function of the UV attenuation for future use. The final section is
a summary of our conclusions.

\section{Radiative transfer through a galactic disc}

In this paper, we use a one-dimensional plane-parallel radiative
transfer model through a galactic disc with clumpy distributions of
stars and dust developed by \cite{ino05}. The clumpiness of the medium
(i.e. the dust distribution) is treated by the mega-grain approximation
which was first proposed by \cite{neu91} and further developed by
\cite{hob93} and \cite{var99}. In this approximation, 
we regard a dusty clump as a huge particle producing
absorption and scattering effects like a normal single dust grain. 
We note that \cite{var99} clearly show the validity of the approximation
by comparisons between the approximate solutions and their
three-dimensional Monte Carlo radiative transfer solutions. The
radiative transfer code of \cite{ino05} can also treat a smooth medium
if we do not use the mega-grain approximation. Since the global geometry
is one-dimensional plane-parallel, we do not consider the bulge and the
radial structure of the disc.

By solving the radiative transfer equations in a single configuration of
stars and dust, we obtain a transmission rate curve, $T_\lambda$, which
is the ratio of the observable and the intrinsic intensities as a function
of the wavelength, $\lambda$. When there are some stellar populations
with different configurations relative to dust in a galactic disc, the
total transmission rate through the disc is
\citep[e.g.,][]{tuf04}
\begin{equation}
  T_\lambda = \sum_i f_{i,\lambda} T_{i,\lambda}\,,
\end{equation}
where $T_{i,\lambda}$ is the transmission rate through the $i$-th
configuration and $f_{i,\lambda}$ is the luminosity weight of the $i$-th
stellar population with a normalization of $\sum_i f_{i,\lambda}=1$. 
Equation (1) means that we can solve the
radiative transfer equations for each stellar population (with each
configuration relative to dust) independently. To obtain the total
transmission rate, we simply sum all the transmission rates with 
luminosity weights (\S2.4).

The transmission rate of each configuration $T_{i,\lambda}$ depends on
\begin{itemize}
 \item the dust model, 
 \item the ISM model, and 
 \item the stellar distribution. 
\end{itemize}
If we divide stars into some populations depending on their age, the
luminosity weight $f_{i,\lambda}$ is determined by 
\begin{itemize}
 \item the age criteria, and 
 \item the star formation history (SFH).
\end{itemize}

In the following, we describe the dust models considered here (\S2.1), 
the ISM model (\S2.2), and the stellar populations and distributions
(\S2.3). Then, we describe how to composite these populations (\S2.4).

\subsection{Dust models}

We consider six dust models in this paper. Table 1 is a summary of the
models and their references. There are two origins of these models; one
is an empirical model by \cite{wg00} (hereafter WG dust) and the other
is a theoretical model by \cite{wei01} and \cite{dra03} (hereafter
Draine dust). There are also four types of dust compositions: the MW, 
the SMC, and the two different LMC types (LMC av and LMC 2). The LMC
av type is an average dust composition over many sight lines toward the
LMC, except for the supershell region around the 30 Doradus, and the LMC
2 type is the dust composition toward the supershell. In the Draine
dust, the bump is assumed to be produced by very small carbonaceous
particles including the ``astronomical'' PAHs designed to fit 
observations \citep{li01}.

\begin{table}
 \caption[]{Dust models}
 \setlength{\tabcolsep}{3pt}
 \footnotesize
 \begin{minipage}{\linewidth}
  \begin{tabular}{lcc}
   \hline
   Model & Reference & $k_{{\rm d},V}$ 
   \footnote{Visual extinction cross section per unit dust mass 
     (cm$^2$ g$^{-1}$).}\\
   \hline
   MW (WG) & \cite{wg00} & $2.60\times10^4$ \\
   SMC (WG) & \cite{wg00} & $1.56\times10^4$ \\
   MW (D) & \cite{dra03} & $2.60\times10^4$ \\
   SMC (D) & \cite{wei01} & $1.56\times10^4$ \\
   LMC av (D) & \cite{wei01} & $1.95\times10^4$ \\
   LMC 2 (D) & \cite{wei01} & $1.89\times10^4$ \\
   \hline
  \end{tabular}
 \end{minipage}
\end{table}%

\begin{figure}
 \centering
 \includegraphics[width=8cm]{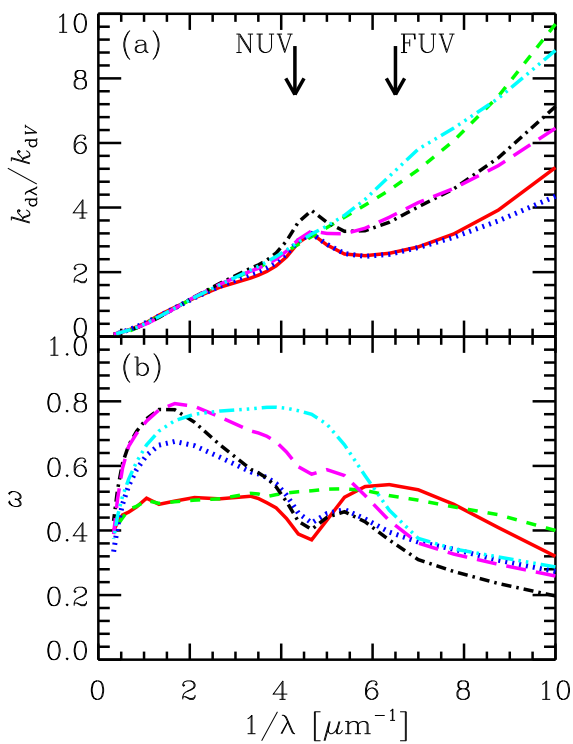}
 \caption{Differences of dust models. The panel (a) shows the extinction
 laws (extinction cross sections normalized by those at the $V$ band) 
 and the panel (b) shows the albedos. The solid and short-dashed lines
 are the MW and the SMC types of Witt \& Gordon (2000),
 respectively. The dotted, dot-dashed, long-dashed, and
 three-dots-dashed lines are the MW, the LMC av, the LMC 2, and
 the SMC types of Draine (2003) and Weingartner \& Draine (2001),
 respectively. The two downward arrows in the panel (a) show the
 effective wavelengths of the two {\it GALEX} filters.}
\end{figure}

\begin{figure}
 \centering
 \includegraphics[width=8cm]{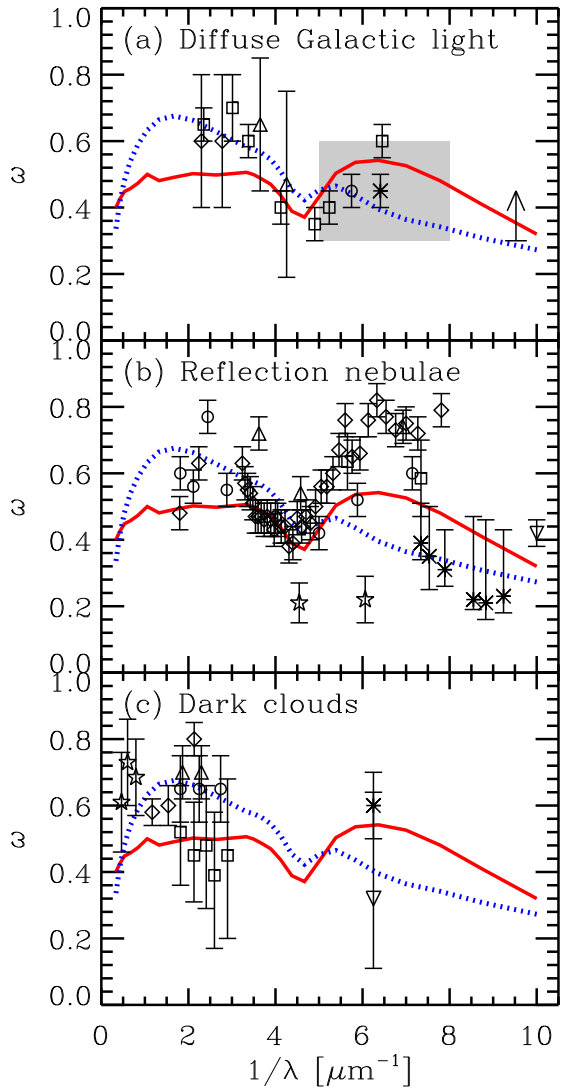}
 \caption{Albedos estimated in the interstellar medium of the Milky
 Way. (a) The diffuse Galactic light; diamonds: Mathis (1973), 
 squares: Lillie \& Witt (1976), triangles: Morgan, Nandy, Thompson
 (1976), arrow (lower limit): Murthy et al.~(1993), shaded area: Murthy
 \& Henry (1995), circle: Witt, Friedmann, \& Sassen (1997), asterisk:
 Schiminovich et al.~(2001). (b) The reflection nebulae; circles: Witt
 et al.~(1982), triangles: Witt et al.~(1992), inverse-triangle: Witt et
 al.~(1993), squares: Gordon et al.~(1994), diamonds: Calzetti et
 al.~(1995), asterisks: Burgh, McCandliss, \& Feldman (2002), stars:
 Gibson \& Nordsieck (2003). (c) The darck clouds; circles: Mattila
 (1970), triangles: Fitzgerald, Stephens, \& Witt (1976), squares:
 Laureijs, Mattile, \& Schnur (1987), diamonds: Witt, Oliveri, \& Schild
 (1990), asterisk: Hurwitz (1994), inverse-triangle: Haikala et
 al.~(1995), stars: Lehtinen \& Mattila (1996). The solid and dotted
 lines are the dust models for the Milky Way by Witt \& Gordon (2000)
 and Draine (2003b), respectively.}
\end{figure}

Fig.~1 shows the extinction cross sections (panel [a]) and the albedos 
(panel [b]) of these models as a function of the wavelength. The
extinction cross sections (i.e. extinction laws) are very similar
between the WG dust and the Draine dust if we compare the same
composition type. We find a prominent bump at 0.22 \micron\ 
($1/\lambda=4.6$ $\micron^{-1}$) in the MW and the LMC av types, 
a weak bump in the LMC 2 type, and no bump in the SMC type. 
We note here that the LMC av type has a rather strong bump 
\citep{fit86,mis99}. 

On the other hand, the albedos are very different between the WG dust
and the Draine dust (panel [b]). Except for the bump region, albedos of
the WG dust (solid and dashed lines) show a flat wavelength dependence in 
$2\,\micron^{-1}<1/\lambda<8\,\micron^{-1}$, whereas those of the Draine
dust (other lines) show a rapid decrease toward shorter wavelengths.
This different wavelength dependence of the albedo significantly affects
the UV colour as shown later (\S3).

In Fig.~2, we show comparisons between albedos of the MW type dusts
and those estimated from observations of the diffuse Galactic light
(panel [a]), of the reflection nebulae (panel [b]), and of the dark
clouds (panel [c]) \citep[see also][]{gor04}. Since the estimated values
show a large dispersion and both dust models are still compatible with
the data, we can not judge which model is better.

\subsection{ISM models}

We consider two cases of the ISM, smooth and clumpy, in a 
plane-parallel disc. We do not consider any systematic vertical
structure of the disc; the mean gas (dust) density is constant
along the vertical axis from the equatorial plane to a height 
$h_{\rm d}$. Above this height, nothing produces absorption and
scattering. For a clumpy medium, we have clumps distributed randomly
in the gas disc, keeping the constant gas density in a volume average.

To model the clumpy medium, we assume a multi-phase ISM picture
\citep[e.g.,][ see \S2 in \citealt{ino05} for details]{fie69}. Assuming
the thermal energy and chemical equilibria in the ISM with temperatures 
lower than $10^4$ K, we have two thermally stable phases (see Fig.~3):
the warm neutral medium (WNM) and the cold neutral medium (CNM). They 
are regarded as the inter-clump medium and clumps, respectively. Based
on \cite{wol03}, we adopt analytical approximations of the relations
between the thermal pressure and the density of these two phases as 
\begin{equation}
  \frac{p/k_{\rm B}}{\rm 10^4\,K\,cm^{-3}} = 
  \frac{n_{\rm H,wnm}}{\rm 1\,cm^{-3}}\hspace{0.5cm}{\rm (WNM)},
\end{equation}
and 
\begin{equation}
  \frac{p/k_{\rm B}}{\rm 10^{4.5}\,K\,cm^{-3}} = 
  \left(\frac{n_{\rm H,cnm}}{\rm 10^3\,cm^{-3}}\right)^{0.7}
  \hspace{0.5cm}{\rm (CNM)}.
\end{equation}
They are shown in Fig.~3 as solid lines. By assuming a mean thermal
pressure, we have a corresponding density contrast between the two
phases. We also assume a mean density of the ISM to obtain a volume
filling factor of clumps. Finally, the clumps are assumed to be
self-gravitating in order to specify their radius (i.e. the Jeans
length).

\begin{figure}
 \includegraphics[width=8cm]{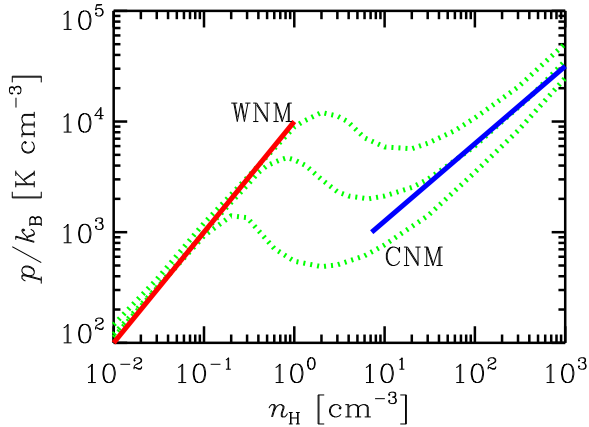}
 \caption{Phase diagram: thermal pressure--hydrogen number density. 
 The dotted curves indicate the thermal equilibrium points in the ISM of
 the MW (top: Galactocentric radius of 3 kpc, middle: 8.5 kpc, and
 bottom: 15 kpc) reproduced from Fig.7 of Wolfire et al.~(2003). The two
 solid lines are the approximate relations of two thermally stable
 phases, the warm neutral medium (WNM) and the cold neutral medium
 (CNM), as expressed in equations (2) and (3).}
\end{figure}

\begin{table}
 \caption[]{Considered physical quantities of the ISM.}
 \setlength{\tabcolsep}{3pt}
 \footnotesize
 \begin{minipage}{\linewidth}
  \begin{tabular}{lcc}
   \hline
   Quantity & Values & Unit\\
   \hline
   $p_{\rm th}/k_{\rm B}$ \footnote{Mean thermal pressure.}
   & $10^{3.0}$, $10^{3.5}$, $10^{4.0}$ & K cm$^{-3}$\\
   $\cal D$ \footnote{Dust-to-gas mass ratio.} 
   & $10^{-3}$, $10^{-2.5}$, $10^{-2}$ & ... \\
   $n_{\rm H}$ \footnote{Mean hydrogen density.}
   & 0.5, 0.75, 1.0, 1.5, 2.0, 3.0, \\
   & 4.0, 6.0, 8.0, 12.0, 16.0, 24.0 & cm$^{-3}$\\
   $h_{\rm d}$ \footnote{Half height of the dusty gas disc.}
   & 50, 100, 150, 200, 250, 300 & pc\\
   \hline
  \end{tabular}
 \end{minipage}
\end{table}%

In order to compare the models with observed galaxies and to extract
the information of dust properties, we cover a wide range of physical
quantities of disc galaxies: the mean ISM hydrogen density
($n_{\rm H}=0.5$--24 cm$^{-3}$), the half height of the dusty disc 
($h_{\rm d}=50$--300 pc), the dust-to-gas mass ratio 
(${\cal D}=0.001$--0.01), and the mean ISM thermal pressure 
($p_{\rm th}/k_{\rm B}=10^{3.0-4.0}$ K cm$^{-3}$), which are
summarized in Table~2. When we keep the two phases described above, 
allowed densities are restricted between the two solid lines in
Fig.~3 for a fixed thermal pressure. Thus, we can take 
$n_{\rm H}=0.5$--6.0 cm$^{-3}$ for $p_{\rm th}/k_{\rm B}=10^{3.0}$ 
K cm$^{-3}$ and $n_{\rm H}=2.0$--24.0 cm$^{-3}$ for 
$p_{\rm th}/k_{\rm B}=10^{4.0}$ K cm$^{-3}$. For 
$p_{\rm th}/k_{\rm B}=10^{3.5}$ K cm$^{-3}$, we can
take all the values of $n_{\rm H}$ listed in Table~2.
Consequently, we have 504 sets of parameters for the clumpy ISM.

In a smooth ISM compared with the clumpy ISM later, 
we do not need to specify the ISM thermal pressure because we do not
make the two-phase medium. 
Hence, we have 216 sets of parameters for the smooth ISM.

As pointed out by \cite{dop05}, the ISM pressure affects the size of H
{\sc ii} regions and the dust temperature in and around these
regions. Since we do not consider the spectral shape of the dust
IR emission in this paper, we omit this effect. In starburst galaxies, a
very high ISM pressure like $p_{\rm th}/k_{\rm B}=10^6$ K cm$^{-3}$ is
observed \citep{lor96,hec99}. However, the two-phase equilibrium is 
not established in such a high ISM pressure as shown in Fig.~3 (there is
no branch of the WNM solution). Since we stand on the two-phase model,
we restrict ourselves within the range of 
$p_{\rm th}/k_{\rm B}=10^{3.0-4.0}$ K cm$^{-3}$ which is observed in the
ISM of MW \citep{mye78}. Thus, our model is for the ``normal'' galaxies
in this respect.

The total disc optical depth at the visual band (0.55 $\micron$) along
the normal of the disc is 
\begin{equation}
  \tau_V = 0.2
  \left(\frac{k_{{\rm d},V}}{10^4\,{\rm cm^2\,g^{-1}}}\right)
  \left(\frac{n_{\rm H}}{1\,{\rm cm^{-3}}}\right)
  \left(\frac{h_{\rm d}}{150\,{\rm pc}}\right)
  \left(\frac{\cal D}{10^{-2}}\right)\,.
\end{equation}
We note that this optical depth is not observable because it is just
proportional to the input dust column density and does not include the
radiative transfer effect. With the visual extinction cross section per
unit dust mass, $k_{{\rm d},V}$, given in Table~1, the considered range
of the input optical depth along the normal axis of the discs is
0.005--25, which corresponds to 0.002--5 $M_{\sun}$ pc$^{-2}$
in terms of the dust column density (see also \S4.2.3),

\subsection{Stellar populations and their distributions}

First, we consider an age-dependent scale-height of the stellar
distribution.  
In the MW, the scale-height of stars younger than 0.1--1 Gyr is $\sim50$
pc, whereas the scale-height of the older stars are $\sim300$ pc
\citep{bin98,rob03}. This is an observational fact although the origin
of this age dependent height is still controversial 
\citep{spi51,tin78,ran91,roc04}. Since the scale-height of the neutral
hydrogen is about 150 pc \citep{bin98}, we have the layering parameters
(ratio of the dusty disc height $h_{\rm d}$ to the stellar scale-height
$h_*$) of $\sim3$ for younger stars and of $\sim0.5$ for older stars. 
We simply set these layering parameters to be constant when the height of
the dusty disc changes in our model.

Next, we divide the younger population into two groups.
Observationally, the youngest stars are often associated with the
molecular clouds. This is natural because stars are formed in the
molecular clouds. On the other hand, this means very inhomogeneous
distribution of the youngest stars. To take into account such a clumpy
stellar distribution, we bury the youngest stars into clumps in the ISM
described in the previous subsection. Eventually, we have three stellar
populations which are called {\it young}, {\it intermediate}, and 
{\it old} stellar populations in this paper (Table~3).

Then, we introduce the age criteria of these three stellar populations. 
The young stellar population is the population embedded in clumps 
(i.e. molecular clouds). Thus, a possible age criterion for this
population is a life-time of molecular clouds, for example, 10 Myr 
\citep[e.g.,][]{bli80}. 
This time-scale is also similar to the life-time of the latest O type
stars. On the other hand, a comparison between the number of the
(ultra-)compact H {\sc ii} regions (i.e. embedded massive stars) and the
number of the visible O stars suggests that only 10--20\% of all O stars
are deeply embedded in molecular clouds \citep{woo89}. To take into
account this fact, we assume a uniform distribution of young stars in a
clump. That is, we have young stars near the surface of the clump as
well as young stars deeply embedded. The former stars would correspond
to visible O stars. The intermediate population has a smaller
scale-height than that of old stars. Thus, the age criterion for this
population is determined by the observed age-dependence of the
scale-height. According to \cite{rob03}, stars with an age of 0.1--1 Gyr
have the smallest scale-height. Here we adopt 300 Myr for the
criterion. This is similar to the time-scale to reach the stationarity
in the UV flux for a continuous star formation.

Table~3 is a summary of the adopted properties of the stellar
populations. To discuss the effect of the clumpiness of the young
population, we consider a smooth distribution for this population in
addition to a clumpy distribution. We also consider two scale heights
(i.e. two layering parameters) for the intermediate population in order
to discuss the effect of the small scale height of this population. 

\begin{table}
 \caption[]{Properties of the stellar populations.}
 \setlength{\tabcolsep}{3pt}
 \footnotesize
 \begin{minipage}{\linewidth}
  \begin{tabular}{lccc}
   \hline
   Population & Age & Layering parameter & Distribution\\
   \hline
   Young & $<10$ Myr & 3.0 & Clumpy/smooth \\
   Intermediate & 10--300 Myr & 3.0/0.5 & Smooth \\
   Old & $>300$ Myr & 0.5 & Smooth \\
   \hline
   \multicolumn{2}{c}{Age of galaxies} & 
   \multicolumn{2}{c}{10 Gyr}\\
   \multicolumn{2}{c}{E-folding time of the SFH} & 
   \multicolumn{2}{c}{5 Gyr}\\
   \hline
  \end{tabular}
 \end{minipage}
\end{table}%

With the age criteria described above, we have luminosity fractions of
the three stellar populations if we assume a SFH and a spectral energy
distribution (SED) of a simple stellar population (SSP). 
Here, we assume three exponentially decaying SFHs with e-folding
time-scales of 1, 3, and 5 Gyr, and a constant SFH. Fig.~4 shows
luminosity fractions for the young stars (panel [a]) and for the
intermediate stars (panel [b]). The age of the galaxy is always assumed
to be 10 Gyr. We have used the SEDs of the SSP with the Solar
metallicity and a Salpeter initial mass function (0.1--100 $M_{\sun}$)
based on the Padova track used in the GRASIL \citep{sil98}. For an
e-folding time larger than about 3 Gyr, differences are small,
especially at a wavelength less than 0.3 \micron. Hereafter, we consider
the case of the e-folding time-scale of 5 Gyr (solid line) as a typical
case.

We comment on a small feature at around 0.3 \micron\ seen especially in
panel (b). This is probably due to the Mg I $\lambda2852$ absorption
line produced in old stars' atmosphere \citep[e.g., ][]{pon98}. 
This line is not prominent in the atmosphere of younger stars, so that
the luminosity fractions of the young stars and the intermediate stars
become relatively large. Our grid of the wavelength has only 25 points
(see Tables C1--C6), and the grid point at $\lambda=0.286$ \micron\ is
affected by the Mg I line. If our wavelength resolution were higher,
more other features would be seen. This point is again discussed in
\S4.1.

\begin{figure}
 \centering
  \includegraphics[width=8cm]{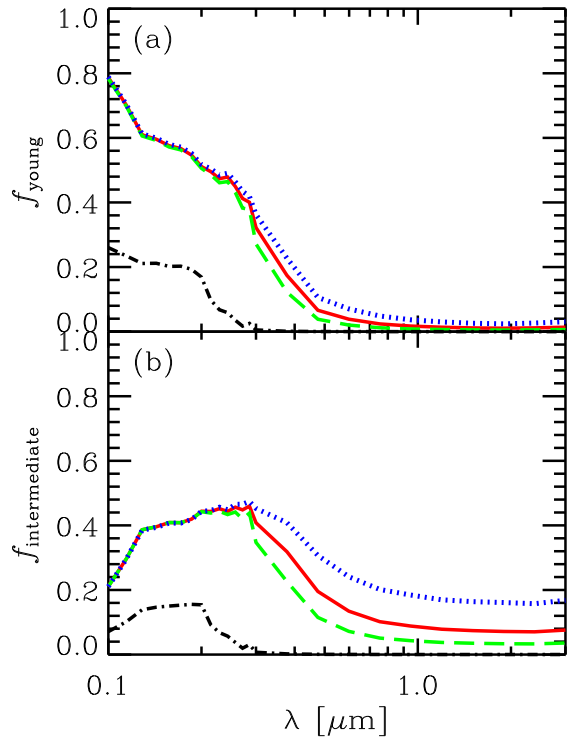}
 \caption{Luminosity fractions of (a) the young and (b) the intermediate 
 stellar populations. The young stars have an age less than 10 Myr,
 and the intermediate stars have an age between 10 Myr and 300 Myr. The
 dotted line is the case of a constant star formation rate. The solid,
 dashed, and dot-dashed lines are the cases of an exponential star
 formation history with a decaying time-scale of 5 Gyr, 3 Gyr, and 1 Gyr,
 respectively. The galactic age of 10 Gyr is assumed for all the cases.}
\end{figure}

\begin{table*}
 \caption[]{Composite models and configurations of dust and stars.}
 \setlength{\tabcolsep}{3pt}
 \footnotesize
 \begin{minipage}{\linewidth}
  \begin{tabular}{lccccc}
   \hline
   Composite model & CCS & CCL & CSS & SSS \\
   \hline
   ISM & Clumpy & Clumpy & Clumpy & Smooth \\
   Young stellar distribution & Clumpy & Clumpy & Smooth & Smooth \\
   Scale height of intermediate stars & Small & Large & Small & Small \\
   \hline
   Composite model & CCS & CCL & CSS & SSS \\
   \hline
   Young & ccl3 & ccl3 & csl3 & ssl3 \\
   Intermediate & csl3 & csl05 & csl3 & ssl3 \\
   Old & csl05 & csl05 & csl05 & ssl05 \\
   \hline
   Configuration & ccl3 & csl3 & csl05 & ssl3 & ssl05 \\
   \hline
   ISM & Clumpy & Clumpy & Clumpy & Smooth & Smooth \\
   Stellar distribution & Clumpy & Smooth & Smooth & Smooth & Smooth \\
   Layering parameter \footnote{Ratio of the height of the dusty disc
   to the scale height of the stellar distribution.}
   & 3.0 & 3.0 & 0.5 & 3.0 & 0.5 \\
   \hline
  \end{tabular}
 \end{minipage}
\end{table*}%

In computations of the radiative transfer, the emissivity decreases
exponentially along the vertical axis with a scale-height $h_*$
for the smooth stellar distribution.
For the clumpy stellar distribution, the emissivity is locally reduced
by multiplying a factor $P_{\rm esc}$ which is the escape energy
fraction from a clump (see eq.[10] in \citealt{ino05}). Here, we use an
advanced version of the model by \cite{ino05}; we take into account the
exponentially decreasing distribution with a scale-height $h_*$ for stars
embedded in clumps. We describe the formulation of this improvement in
appendix A. Above the disc height $h_{\rm d}$, there is no stars for the
clumpy case because of the absence of clumps.

\subsection{Composite of star/dust configurations}

Now, we composite the three stellar populations introduced in the
previous subsection. This is done by equation (1). 
We call the dust/star geometry for a single stellar population 
``configuration'', and call a combination of three stellar 
populations with different configurations ``composite model''. 
Considering composite models, we have three
points to choose the set-up: the clumpy/smooth ISM, the clumpy/smooth
young stellar distribution, and the small/large scale height for the
intermediate stellar population. Here, we consider four composite models 
listed in the top part of Table~4. 
First, as a standard model, we consider the case with the clumpy ISM, 
the clumpy young stellar distribution, and the small scale height of the
intermediate stellar population. This case is called ``CCS''. 
To examine the effect of the scale height of the intermediate
population, we consider the case with the clumpy ISM, the clumpy young
stellar distribution, and the large scale height of the intermediate
population. This is called ``CCL''. To assess the effect
of the young stellar distribution, we consider the case with the clumpy
ISM, the smooth distribution of the young stars, and the small scale
height of the intermediate stars. This is called ``CSS''. 
Finally, we consider the case with the smooth ISM, the smooth distribution
of the young stars, and the small scale height of the intermediate stars, 
which is useful to discuss the effect of the ISM clumpiness and is called
``SSS''.

To produce these four composite models, we need five configurations of
dust and stars. The bottom of Table~4 is a summary of these
configurations.
The first case is the clumpy ISM, the clumpy stellar distribution, and
the layering parameter of 3.0 (i.e. smaller scale height of the stellar
distribution than the height of the dusty disc). This is called ``ccl3'' 
and for the young stellar population in the CCS and the CCL models. 
The next is the clumpy ISM, the smooth stellar distribution, and the
layering parameter of 3.0. This is called ``csl3'' and for the
intermediate population in the CCS model and the young and intermediate
populations in the CSS model. The third is the clumpy ISM, the smooth
stellar distribution, and the layering parameter of 0.5 (i.e. larger
scale height of the stellar distribution than the height of the dusty
disc). This is called ``csl05'' and for the intermediate and old
populations in the CCL model and the old stellar population in the CCS
and CSS models. The fourth is the smooth ISM, the smooth stellar
distribution, and the layering parameter of 3.0. This is called ``ssl3''
and for the young and intermediate populations in the SSS model. The
last one is the smooth ISM, the smooth stellar distribution, and the
layering parameter of 0.5. This is called ``ssl05'' and for the old
stellar population in the SSS model. 
The middle of Table~4 shows correspondences between
the configurations and the stellar populations in the composite models.
\footnote{For each configuration, we consider 6 dust models (\S2.1)
and 504 sets of physical quantities in a clumpy ISM or 216 sets in a
smooth ISM (\S2.2). In total, we have 3,024 cases for each csl05, csl3,
and ccl3, and 1,296 cases for each ssl05 and ssl3. After solving the
radiative transfer equation with the number of the angular coordinate of
16, we have 48,384 transmission rate curves ($T_\lambda$, transmission
rates as a function of the wavelength whose resolution of 25 from 0.1
\micron\ to 3.0 \micron) for each csl05, csl3, and ccl3, and 20,736 for
each ssl05 and ssl3. Finally, we composite $T_\lambda$ of these
configurations with luminosity weights as shown in Fig.~4 and obtain 
48,384 composite $T_\lambda$ for each CCS, CCL, and CSS cases, 
and 20,736 for the SSS case. In total, we have 165,888 transmission rate
curves for the four composite models.}

\section{{\it GALEX} colours and dust properties}

In this section, we show that the {\it GALEX} colour is strongly
affected by dust models, especially by the presence of the
bump and by the wavelength dependence of the albedo, rather than 
by the geometry in the radiative transfer. 
First, differences among the configurations (\S3.1) and the composite
models (\S3.2) are shown, and then, the suitable dust models for the
galaxies observed with the {\it GALEX} are discussed (\S3.3).
We note here that the filter transmission efficiencies of the {\it
GALEX} two band passes are correctly taken into account in all
calculations.

\subsection{Comparison of configurations}

Here, we compare the UV dust attenuations among different configurations
(the bottom part of Table~4) and dust models.
To concentrate on effects of dust models and configurations, 
we consider the colour excess of the two {\it GALEX} bands, 
$E(FUV-NUV)=A_{FUV}-A_{NUV}$. Its dependence on the assumed SED is very
weak (we assume a flat SED, $f_\lambda \propto \lambda^0$, for  
simplicity in this and the next subsection). Fig.~5 shows the FUV
attenuation, $A_{FUV}$, as a function of $E(FUV-NUV)$. The panels (a) to
(f) correspond to the six dust models described in \S2.1. In each panel, 
the solid line is the locus expected from the extinction law
(i.e. distant uniform screen geometry) and the dashed line is that from 
the Calzetti law. When making the figure, we calculated $A_{FUV}$ and 
$A_{NUV}$ (and then $E(FUV-NUV)$) from the transmission rate curves
obtained from the radiative transfer calculations with each
configuration. Thus, we had 48,384 (8,064 $\times$ 6 dusts) points each
for csl05, csl3, and ccl3 (clumpy ISM cases) and 20,736 (3,456 $\times$
6 dusts) points each for ssl05 and ssl3 (smooth ISM cases) on the
$E(FUV-NUV)$--$A_{FUV}$ plane. Then, we divided these points into
several bins in $A_{FUV}$ and obtained the maximum, minimum, and median
values of the distribution of $E(FUV-NUV)$ for each bin, each dust, and
each configuration. The plotted symbols indicate the median location for
five configurations: diamonds, squares, open circles, triangles, and
filled circles are ssl05, ssl3, csl05, csl3, and ccl3, respectively. The
bin widths are shown as the vertical error-bars, and the full
distribution ranges of $E(FUV-NUV)$ are shown as the horizontal
error-bars. 

\begin{figure*}
 \centering
 \includegraphics[width=14cm]{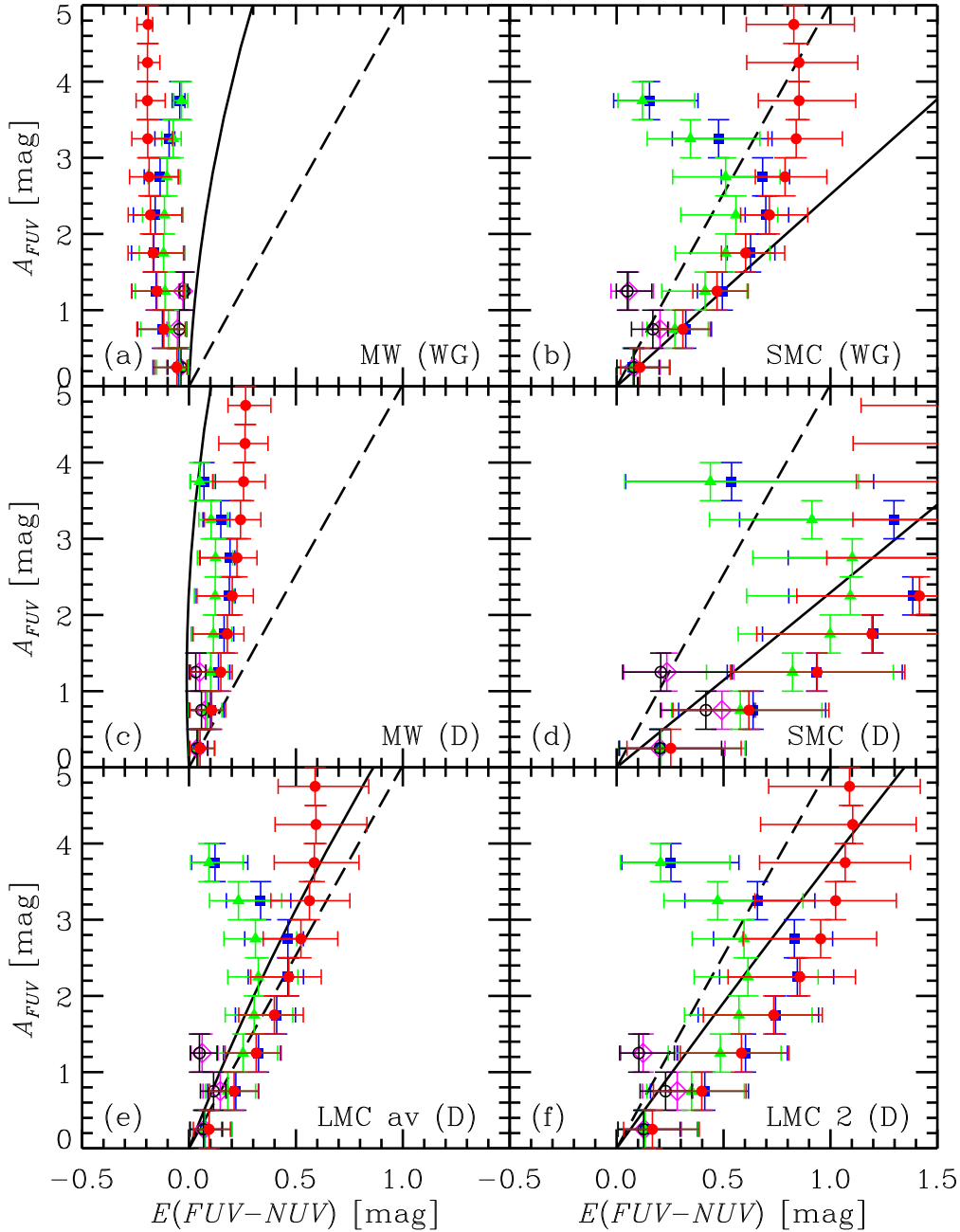}
 \caption{$FUV$ attenuation $A_{FUV}$ as a function of the colour excess
 $E(FUV-NUV)$. The solid lines are the relation expected from the
 extinction law (i.e. distant uniform screen). The dashed lines are the
 case of the Calzetti law. The open diamonds, filled squares, open
 circles, filled triangles, and filled circles are configurations of
 ssl05, ssl3, csl05, csl3, and ccl3, respectively (see Table~4). The
 vertical error-bars show the bin widths, the horizontal error-bars show 
 the full widths of the distribution of $E(FUV-NUV)$ in each bin, and the
 symbols show the median locations of the distribution.}
\end{figure*}

Through all the panels, we see a large variation of the plotted points
depending on the dust models. From comparisons between the MW and SMC
types ([a] and [b], or [c] and [d]), we find the effect of the bump is
very large; the most of the MW cases distribute within 
$|E(FUV-NUV)| \la 0.3$ mag because of the bump, whereas the SMC cases
can reach at $E(FUV-NUV) \ga 0.5$ mag. Indeed, we see {\it ``blueing''}
due to the bump in the MW (WG) case (panel [a]). The LMC cases ([e] and
[f]) are located between the MW and the SMC cases. Thus, we observe the
trend that the UV colour becomes redder as the bump becomes weaker 
\citep[e.g.,][]{wg00}.

We newly find that the wavelength dependence of the albedo also strongly
affects the UV colour. From comparisons between two dust models with the
same extinction law ([a] and [c], or [b] and [d]), we find that the
Draine dust cases are systematically redder than the WG dust cases. 
Indeed, the MW (WG) case shows blueing, whereas the MW (D) case shows a
slight reddening. While the most of the points of the WG dust appear in
the left-hand side (bluer colour) of the extinction law (solid line),
those of the Draine dust are in the right-hand side (redder colour) of
the extinction law, except for some large $A_{FUV}$ cases. This
difference is caused by the difference in albedos. As shown in Fig.~1
(b), the albedos of the Draine dusts decrease rapidly from optical to UV. 
On the other hand, those of the WG dusts are almost flat, except for the
bump domain. Relative to the WG dusts, the Draine dusts easily absorb
the FUV photons and scattered out the NUV photons, so that the 
{\it GALEX} colour becomes redder, i.e. $E(FUV-NUV)$ becomes larger.

In each panel, we compare the different configurations. All
configurations, except for the ccl3, show a maximum in $A_{FUV}$ and a
turnover of $E(FUV-NUV)$ in the middle of their locus. At the largest
$A_{FUV}$, the minimum $E(FUV-NUV)$ (or maximum for the MW [WG] case)
reaches 0 mag, i.e. no reddening/blueing.\footnote{The displacements of
the symbol location from $E(FUV-NUV)=0$ at the largest $A_{FUV}$ in
Fig.~5 are due to the binning effect. Note that the symbol location is
median of the $E(FUV-NUV)$ distribution in each bin.}  Generally, the
observed intensity along a ray is determined by the sum of two
intensities along the ray: the intensity transmitted through the dusty
disc and the intensity from the source outside the dusty disc (see
eq.[21] in \citealt{ino05}). If the disc opacity increases, the first
intensity decreases, and the observed intensity is dominated by the
second intensity. In this case, the transmission rate (the ratio of the
observed to intrinsic intensities) reaches an asymptotic value depending
on the relative fraction of the second intensity in the intrinsic total
intensity, i.e., the relative amount of the source outside the dusty
disc. The maximum
$A_{FUV}$ is originated from the asymptotic value. Only the layering
parameter determines the amount of the source outside the disc when we
consider a single configuration. Thus, the wavelength dependence of the
transmission rate disappears (i.e. a grey attenuation), and then, 
we have $E(FUV-NUV)=0$ at the maximum $A_{FUV}$. 
The ccl3 case has no source outside the dusty disc because there is no
clump outside the disc. Thus, this case does not have the maximum
$A_{FUV}$ and shows only a weak turnover outside the plotted region.

The clumpy ISM cases without the embedded stars (csl3 and csl05) are 
$\sim 0.1$--0.3 mag bluer (redder for the MW [WG] case) than the smooth
ISM cases (ssl3 and ssl05). This is because a clumpy ISM is less opaque
than a smooth ISM. However, the case with the embedded stars (ccl3) has
an additional local opacity due to clumps which redden (blue for the
MW[WG] case) the UV colour, so that its colour excess coincides with the 
smooth ISM case (ssl3) before the turnover in $E(FUV-NUV)$ (a small
$A_{FUV}$) and becomes redder (bluer for the MW [WG] case) than the
smooth case after the turnover (a large $A_{FUV}$).

In summary, the presence of the bump and the UV wavelength dependence of
the albedo strongly affect the UV colour. Except for the case with the
embedded stars, the layering parameter puts the maximum UV attenuation
at which the colour excess becomes zero. There is also the maximum (or
minimum for the blueing case) colour excess. The effects of the ISM
clumpiness and the stellar clumpiness are not very large before the
maximum colour excess.

\subsection{Comparison of composite models}

Here, we compare the {\it GALEX} colour excess, $E(FUV-NUV)$, expected
from the composite models summarized in the top part of Table~4.
Fig.~6 is made by the same way as Fig.~5, and 
shows the relation between $E(FUV-NUV)$ and $A_{FUV}$. 
Each composite model is shown by each own symbol: SSS (squares), CSS
(triangles), CCL (open circles), and CCS (filled circles). Six panels of
(a) to (f) show six dust models considered here.

\begin{figure*}
 \centering
 \includegraphics[width=14cm]{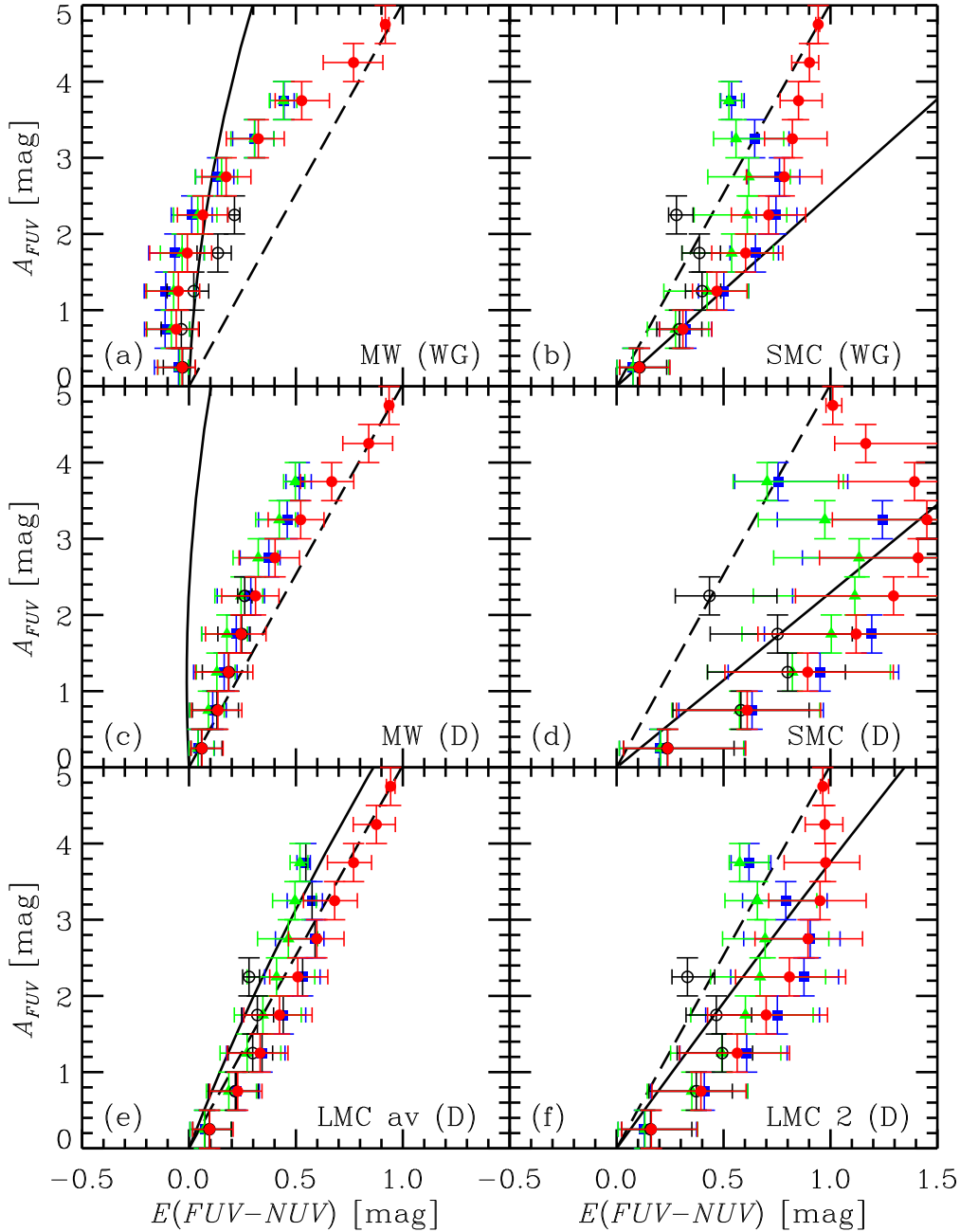}
 \caption{Same as Fig.~5, but composite cases. The filled squares, filled
 triangles, open circles, and filled circles are the SSS, CSS, CCL, and CCS
 models (see Table~4), respectively.}
\end{figure*}

\begin{figure*}
 \centering
 \includegraphics[width=14cm]{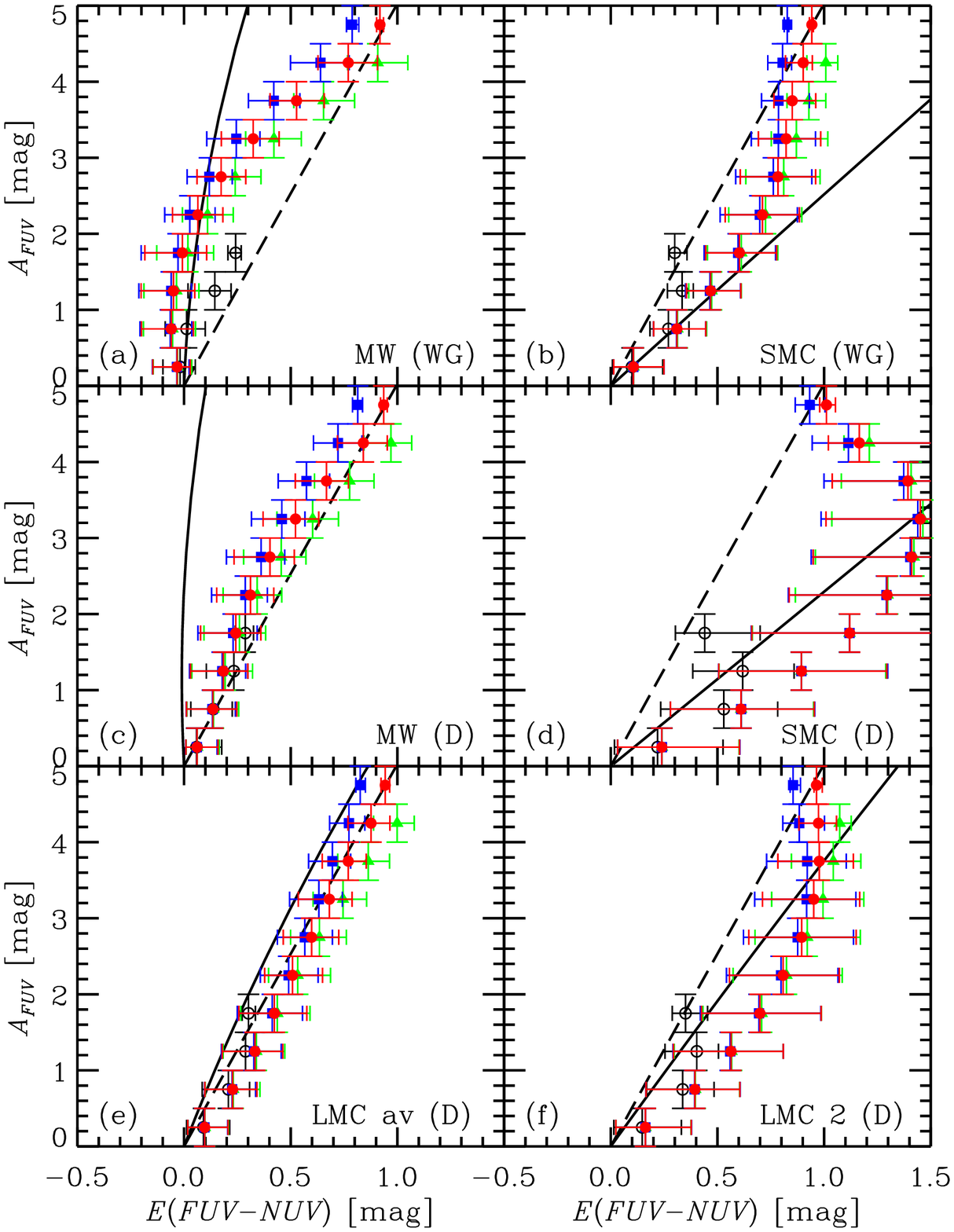}
 \caption{Same as Fig.~5, but the CCS model (Table 4) with different
 star formation histories. The open circles, filled triangles, and 
 filled circles are the cases with e-folding time-scales of 1 Gyr, 3
 Gyr, and 5 Gyr, respectively. The filled squares are the cases with a
 constant star formation rate.}
\end{figure*}

As found in Fig.~5, we find that the effect of the bump is large
from comparisons between panels ([a] and [b]) or ([c] and [d]) and that
the wavelength dependence of the albedo is also important from
comparisons between panels ([a] and [c]) or ([b] and [d]), particularly
for $A_{FUV}\la3$ mag. Indeed, $E(FUV-NUV)$ of the MW types ([a] and [c])
is about 0.5--1.0 mag bluer than that of the SMC types ([b] and
[d]). This is the effect of the bump. On the other hand, 
$E(FUV-NUV)$ of the Draine dusts ([c] and [d]) is about 0.2--0.5 mag
redder than that of the WG dusts ([a] and [b]). This is the effect of
the albedo. Thus, the rapidly decreasing albedo toward shorter 
wavelengths like the Draine dust can partially compensate the blueing
effect of the bump.

As seen in Fig.~5, we observe a turnover in $E(FUV-NUV)$ for some
cases. However, the final points (the largest $A_{FUV}$) in each
composite model do not reach $E(FUV-NUV)=0$. This shift from
$E(FUV-NUV)=0$ is caused by the composite process, 
i.e. the effect of the age-dependent attenuation. 
This effect can be understood by equation (1). If the disc opacity is
large enough, the attenuations for any populations become grey; 
$T_{i,\lambda}$ loses the wavelength dependence. Even in this case, 
we still have the wavelength dependence of $f_{i,\lambda}$. Thus, 
$E(FUV-NUV)$  for an opaque disc is determined by the wavelength
dependence of the luminosity weight, in other words, the SFH and the age
criteria of the stellar populations. More interestingly, the attenuation
law becomes independent of the dust model in such a case. 
Indeed, we find that the locations of the largest $A_{FUV}$ point of 
each composite model in six panels (i.e. six dust models) are very
similar to each other. This causes rather large $E(FUV-NUV)$ even for
the MW dust (panels [a] and [c]). We can expect a red {\it GALEX} colour
even with the bump \citep[see also][]{pan05}. 
The age-dependent attenuation reduces the effect of the bump through the
wavelength dependence of the luminosity weights.

In each panel, when we compare the CCS and CCL models, the effect of the
intermediate population can be understood. With a large scale-height of
the intermediate population (CCL), we can not reach $A_{FUV}\ga2.5$ mag,
whereas we reach $A_{FUV}\sim5$ mag with a small scale-height of the
population. As shown by \cite{bua05}, 
FUV attenuations of many nearby galaxies selected by FIR are more than
2.5 mag. Thus, an intermediate population with a small scale-height is
likely to be required. When we compare the
CCS and CSS or SSS models, we find the effect of the clumpy young
stars. Because of the additional local opacity due to clumps, the CCS
model reaches $\sim1$ mag larger $A_{FUV}$ than the CSS and SSS models. 
We find that the effect of the ISM clumpiness is small for $A_{FUV}\la3$
mag from a comparison between the CSS and SSS models. In the following
discussions, we deal with only the CCS model which consists of the young
population embedded in clumps, the intermediate population with a
small-scale height, and the old population with a large-scale height,
i.e. the most realistic model in this paper.

Fig.~7 shows the effect of the SFH on the $E(FUV-NUV)$--$A_{FUV}$
relation for the CCS model. For an e-folding time-scale longer than 3
Gyr, the difference in $E(FUV-NUV)$ is almost zero for $A_{FUV}\la2$
mag, and small shifts ($\sim0.1$ mag) of the largest $A_{FUV}$ are
observed. This small difference is due to the
small difference of the luminosity weights in the UV ($\lambda \la 0.3$
\micron) as shown in Fig.~4. On the other hand, a short e-folding
time-scale like 1 Gyr produces a difference on the
$E(FUV-NUV)$--$A_{FUV}$ plane. This case reaches only $A_{FUV}\la2$ mag
because the radiation is dominated by the old population emitting from
the outside of the dusty disc.

In summary, the effect on the {\it GALEX} colour significantly depends
on the dust model (with/without the bump and the albedo) for a small
UV attenuation, whereas it becomes independent of the dust model for the
largest UV attenuation. The clumpy young stars and the intermediate
population with a small scale-height increase the UV attenuation. The
effect of the intermediate population is larger than that of the young
population, at least in the SFH considered here. On the other hand, the
ISM clumpiness has a relatively small effect on the UV colour
excess. The SFH has a very small effect on the UV colour excess if we
assume a smooth SFH with an e-folding time-scale larger than 3 Gyr.

\subsection{IR-to-UV flux ratio and {\it GALEX} colour}

\begin{figure*}
 \centering
 \includegraphics[width=14cm]{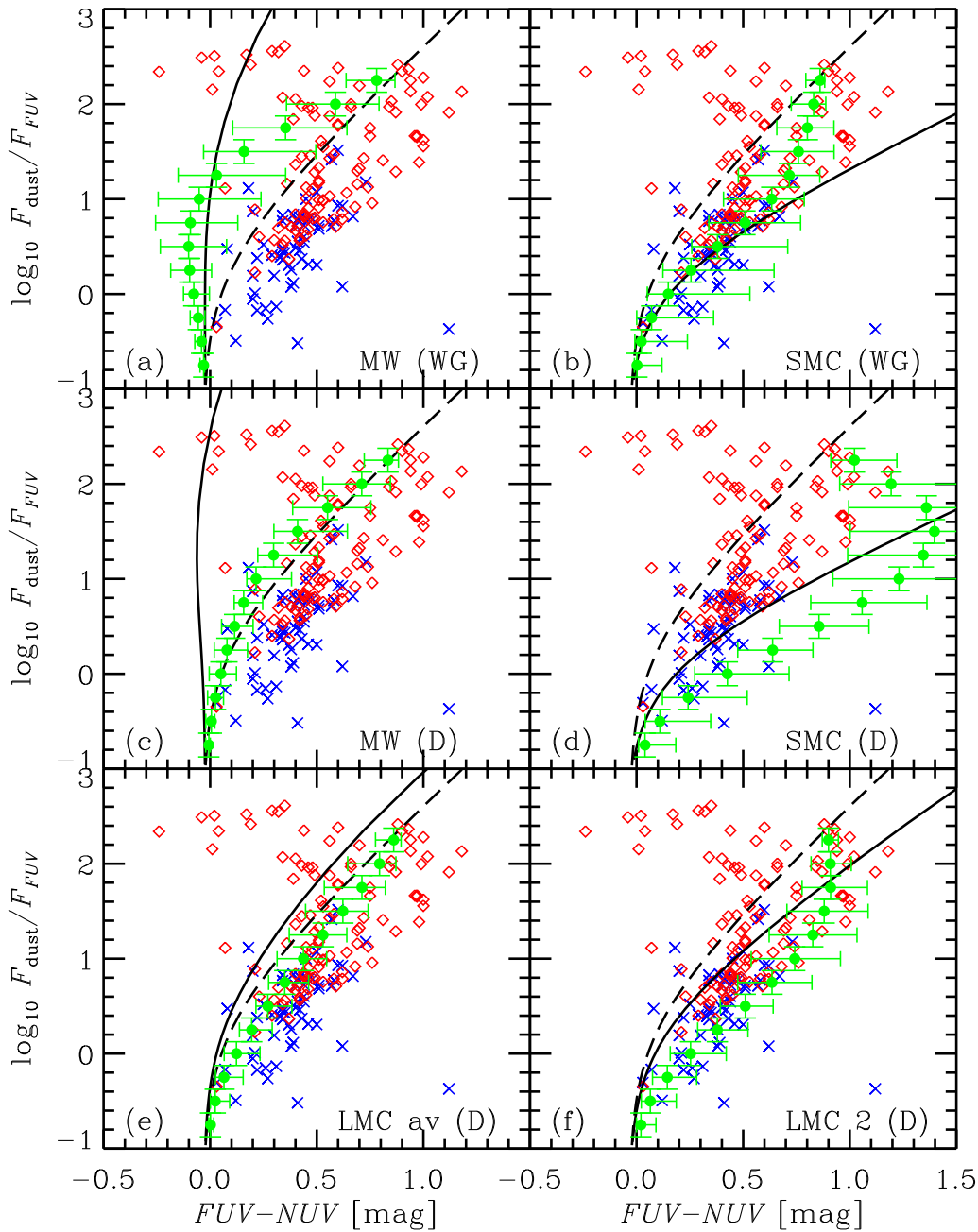}
 \caption{Dust IR-to-FUV flux ratio and the {\it GALEX} colour. The
 crosses and diamonds are observed data of the NUV selected and the FIR
 selected nearby galaxies, respectively, taken from Buat et al.~(2005)
 (see also \citealt{igl05}). 
 The solid and dashed lines correspond to the extinction law and the
 Calzetti law, respectively. We show the CCS model (Table~4). The points
 predicted from the model are divided into several bins in the flux
 ratio. The vertical error-bars are the bin width, the horizontal
 error-bars are the full width of the distribution of $FUV-NUV$ in each
 bin, and the symbols are the median of the distribution.}
\end{figure*}

As shown in the previous subsection, the {\it GALEX} colour is very
sensitive to the presence of the bump and the wavelength dependence
of the albedo. Conversely, we may assess the dust models by comparing
with the observed {\it GALEX} colours. Fig.~8 shows the diagram
of the IRX (dust IR-to-UV flux ratio, $F_{\rm dust}/F_{FUV}$) and the
{\it GALEX} colour, $FUV-NUV$. The crosses and diamonds are the observed
data of the nearby galaxies selected by NUV and FIR, respectively, taken
from \cite{bua05} (see also \citealt{igl05}). The filled circles and
error-bars are made by the same procedure done in Fig.~5. 
We show only the CCS model with the e-folding time-scale of 5 Gyr for
the comparison. We calculated the model dust IR flux as the total
absorbed flux. This procedure is described in appendix B. The UV flux is
defined as the UV flux density multiplied by the effective wavelength of
the UV filter. The solid and dashed lines in Fig.~8 are the loci
expected from the extinction law and the Calzetti law, respectively.

By comparing the model points with the observed data, we assess which
dust model is suitable for the observed galaxies. 
The SMC (WG) case shows a very good agreement with the data of the UV
selected galaxies and the FIR selected galaxies with 
$F_{\rm dust}/F_{FUV}\la100$ (panel [b]). The LMC av (D) and the LMC 2
(D) cases are also compatible with the data (panels [e] and [f]). The
colours predicted by the MW (D) case are still $\sim0.2$--0.3 mag bluer
than the observed ones because of a strong bump and a shallow UV slope
in the extinction law (panel [c]), although a different IMF like a
Kroupa IMF could reduce the discrepancy 
\citep{pan05}.\footnote{\cite{pan05} have reproduced the data of the NUV
selected galaxies with a Draine's MW dust (but a different version from
this paper) better than here. This may be due to some differences
between the two papers; they treat a 2-D disc + bulge with a smooth
medium, whereas we treat a 1-D plane-parallel disc with a clumpy
medium. Moreover, a different age criterion of the intermediate
population may play a role; \cite{pan05} show that a shorter criterion
gives a redder colour.} 
On the other hand, the predicted colours of the MW (WG) case are largely
separated from the observed data, say $\sim0.5$ mag at  
$F_{\rm dust}/F_{FUV}\sim10$ (panel [a]). For the SMC (D) case (panel
[d]), the predicted colours are too red ($\sim0.5$ mag at  
$F_{\rm dust}/F_{FUV}\sim10$) because of a rapid decrease of the albedo
between the two {\it GALEX} bands as shown in Fig.~1 (b).

Interestingly, there is no model which reproduces the FIR selected
galaxies with $F_{\rm dust}/F_{FUV}\ga100$ where ULIRGs also
distribute \citep{gol02}.\footnote{Only a few galaxies in the FIR
selected sample of \cite{bua05} are ULIRGs.} 
As discussed in \S3.2, for an opaque disc, we expect to have an 
attenuation law (i.e. transmission rate curve) independent of 
dust properties. In fact, we find that the locations of the most opaque
point in each panel are very similar; all dust models predict a very
similar position on the diagram for $F_{\rm dust}/F_{FUV}\ga100$. 
However, the real galaxies show a very large dispersion in the region. 

\cite{bur05b} could not explain the same galaxies by their analysis
either and suggested that an effect of ``decoupling'' is important; the
stellar populations producing the UV and the IR are completely
different. For example, the UV radiation comes from the population
outside the obscured region, whereas the population heating dust
which emits the IR radiation is embedded there. In this case, the UV
colour is decoupled with the UV attenuation traced by 
$F_{\rm dust}/F_{FUV}$.\footnote{We can still rely on the IR-to-UV flux
ratio for an indicator of the {\it total} UV attenuation because the
flux ratio is the ratio of the absorbed to observed radiation energies 
(escaped from the obscured region + emitted from the outside population)
based on the energy conservation.}
In the framework of this paper, such a ``decoupling'' would take place 
if we consider an intermittent SFH. Now, we have three stellar
populations: young and intermediate ones embedded in clumps and in the
dusty disc, and old one distributed diffusely to the outside of the
disc. Under an intermittent SFH with a time-scale longer than $\sim300$
Myr (age threshold between the intermediate and old populations), we can
expect that the luminosity weights strongly vary along the time, and
then, the position of the most opaque case on the IRX--UV colour diagram
would vary. 

In summary, the {\it GALEX} data are consistent with WG's SMC dust
and Draine's dusts with the bump (LMC av, LMC 2, and possibly MW). 
If we consider that the carrier of the UIR very common among the nearby
galaxies \citep[e.g.,][]{gen00} is also the bump carrier as suggested by
some laboratory investigations \citep[e.g.,][]{sak83}, dust with the
bump is more favorable. For a flat wavelength dependence of the albedo
like the WG dust, a strong bump seen in the MW makes the {\it GALEX}
colour too blue. However, a weak bump like the LMC 2 type could be
still compatible with the {\it GALEX} data.
Some FIR selected galaxies which have a large IR-to-UV flux ratio 
($F_{\rm dust}/F_{FUV}\ga100$) cannot be explained by a smooth SFH
model, may suggesting their recent episodic star-formation.

\section{Mean attenuation laws}

From the calculated transmission rate curves, we derive mean attenuation
laws which would be useful for correcting the observed data of
galaxies for the dust attenuation and for predicting the observable SEDs
in theoretical works. We show only the CCS model (Table 4) in this
section. 

\subsection{Mean UV-to-NIR attenuation laws}

\begin{figure*}
 \centering
 \includegraphics[width=14cm]{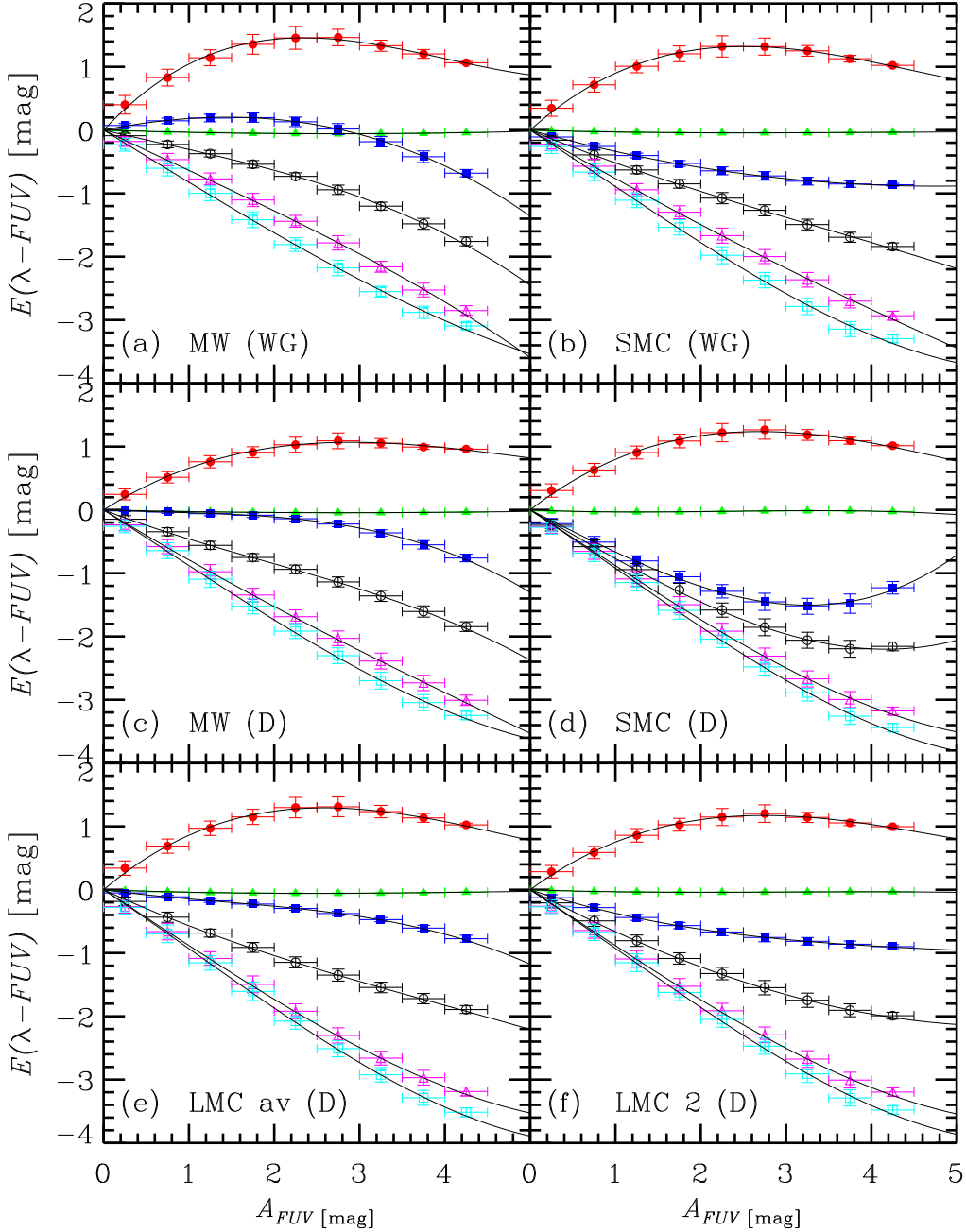}
 \caption{Colour excesses relative to the FUV at various wavelengths,
 $E(\lambda-FUV)$, as a function of the FUV attenuation, $A_{FUV}$. The
 filled circles, filled triangles, filled squares, open circles, open
 triangles, and open squares are the colour excesses at 
 $\lambda=0.100$, 0.157, 0.229, 0.300, 0.599, and 1.19 \micron,
 respectively. The horizontal error-bars indicate the $A_{FUV}$ bin
 width. The plotted points and the vertical error-bars are the mean and
 the standard deviation of the $E(\lambda-FUV)$ distribution in each
 bin. The thin solid curves are the 3rd order polynomial fits.}
\end{figure*}

\begin{figure*}
 \centering
 \includegraphics[width=14cm]{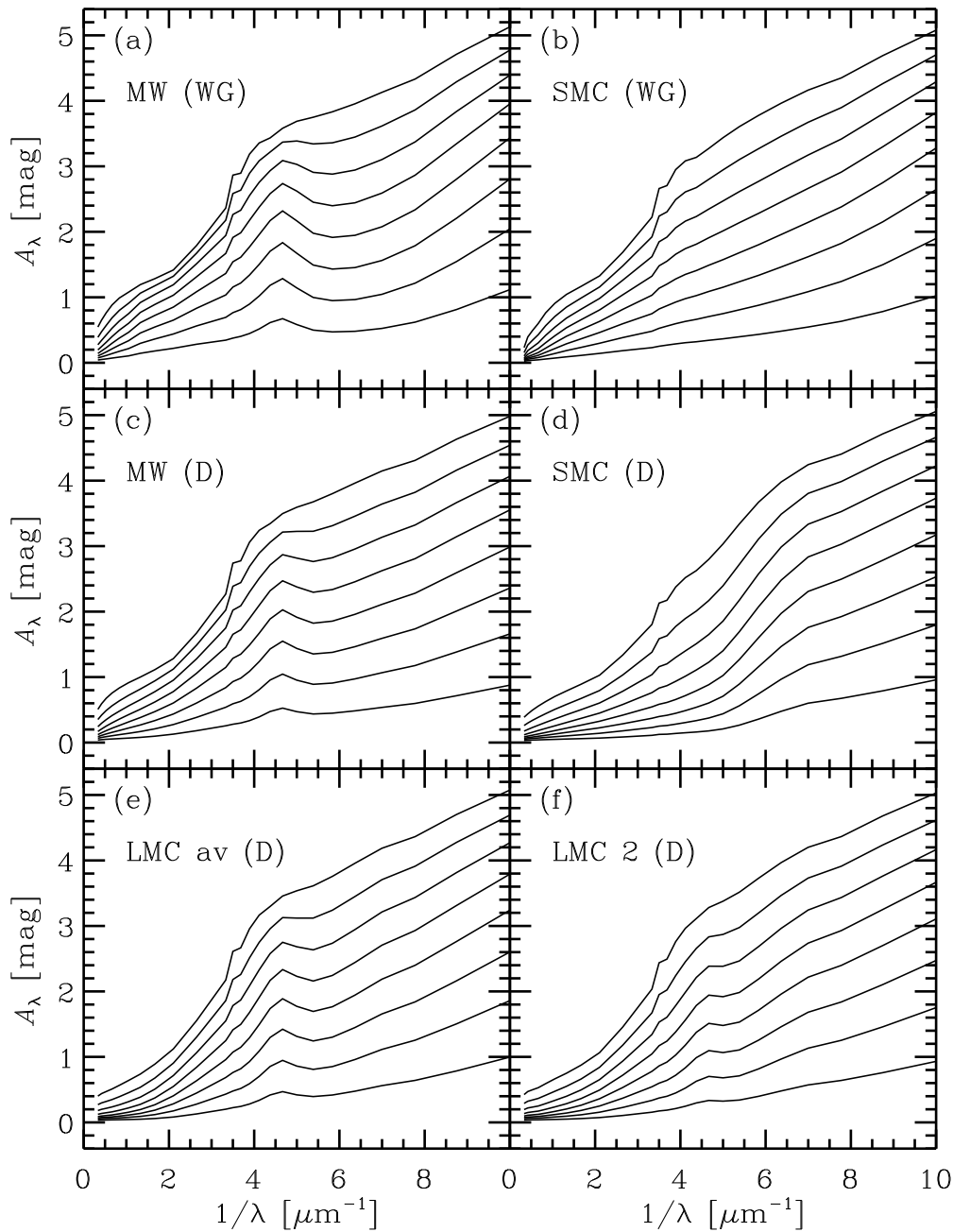}
 \caption{Mean attenuation laws. In each panel, eight attenuation laws
 are shown for $A_{FUV}=0.5$, 1.0, 1.5, 2.0, 2.5, 3.0, 3.5, and 4.0 mag
 from the bottom to the top.}
\end{figure*}

We find that the FUV attenuation, $A_{FUV}$, represents the
global shape of the attenuation laws. In Fig.~9, the
colour excesses relative to $A_{FUV}$ at various wavelengths,
$E(\lambda-FUV)=A_\lambda-A_{FUV}$, are shown as a function of
$A_{FUV}$. To make the figure, we first averaged the transmission rates, 
$T_\lambda$, over the angle between a ray and the disc normal
(i.e. inclination angle) via equation (B2). Then, we divided the
inclination averaged $T_\lambda$ into some bins in $A_{FUV}$ and
calculated the mean and the standard deviation in the distribution of 
$E(\lambda-FUV)$ in each bin and each dust model. 
The location of symbols and the vertical error-bars in
Fig.~9 show the mean (not the median) and the standard deviation (not
the full width), respectively. The horizontal error-bars indicate the
bin widths. Fig.~9 shows that $E(\lambda-FUV)$ can be expressed as a
function of $A_{FUV}$ very nicely. Indeed, a typical standard deviation
is as small as 0.1 mag. Thus, we tried to fit the calculated
$E(\lambda-FUV)$ by a 3rd order polynomial function of $A_{FUV}$ for
each dust model as 
\begin{equation}
 E(\lambda-FUV) = \alpha(\lambda) A_{FUV} + \beta(\lambda) {A_{FUV}}^2 
  + \gamma(\lambda) {A_{FUV}}^3\,.
\end{equation}
Note that we have assumed $E(\lambda-FUV)=0$ when $A_{FUV}=0$.
The order of the polynomial function was determined, based on the
Akaike's Information Criterion \citep{tak00b,tak00}; we tried to fit up
to 6th order and confirmed that 3rd order is enough. The resulting
parameters for each wavelength and each dust model are tabulated in
appendix C. In Fig.~9, the fitting results are shown as thin solid
curves.

Fig.~10 shows the mean attenuation laws obtained by the polynomial
fit of $E(\lambda-FUV)$. We observe that the significance of the
bump reduces as $A_{FUV}$ increases (panels [a], [c], [e], and [f]). 
We also find a different slope in the UV ($1/\lambda\sim5$--7
\micron$^{-1}$) between the two SMC cases (panels [b] and [d]); a
steeper rise of the SMC (D) case is due to a rapid decline of its
albedo (Fig.~1 [b]). We also note that the shape of the attenuation
law for the most opaque case (top curves, $A_{FUV}=4$ mag) in each panel
is very similar to those in other panels as discussed in \S3.2.

We comment on the small feature seen at $1/\lambda \simeq 3.5$
\micron$^{-1}$ in all panels of Fig.~10. This is due to the small feature
seen in Fig.~4, the stellar Mg I $\lambda2852$ absorption line. This
line is prominent in old stars' atmosphere but not in young and
intermediate stars' one. Such an age-dependent strength of the line
appears as a small feature in the luminosity fractions of young and
intermediate stars. As discussed in \S3.2, the wavelength dependence of
the attenuation law (i.e. transmission rate $T_\lambda$) is determined
by the luminosity fractions if the disc is enough opaque. Thus, such a
stellar feature appears in the attenuation law of an opaque
disc. Indeed, the feature in Fig.~10 is clearer for larger $A_{FUV}$
cases. Interestingly, an age-dependent stellar feature can appear in the
attenuation law through an age-selective attenuation.

To quantify the bump reduction and the slope change along the FUV
attenuation seen in Fig.~10, we tried to fit the normalized mean
attenuation laws by a power law plus a Gaussian bump with the same way
as \cite{bur05b}: 
\begin{equation}
 \frac{A_\lambda}{A_{\lambda_{\rm b}}} = 
  (1-B) \left(\frac{\lambda}{\lambda_{\rm b}}\right)^{-p}
  + B \exp \left(-\frac{\lambda-\lambda_{\rm b}}{\sigma}\right)^2\,,
\end{equation}
where the bump position $\lambda_{\rm b}=0.2175$ \micron\ and the bump
width $\sigma=0.02$ \micron. Although this function does not give a very
good fit, we adopt it for simplicity. We left two free parameters: the
power law index $p$ and the bump amplitude $B$. Fig.~11 shows results of
the fit. We find clearly that the global slope of the attenuation law
reduces as $A_{FUV}$ increases (panel [a]). This is consistent with
previous investigations \citep[e.g.,][]{var99,fer99,wg00,pie04,pan05}
For each dust model, the power law index decreases from a larger value
than its extinction law (i.e. steeper attenuation) to a smaller value
(i.e. greyer attenuation), and finally, the indices converge a value
similar to that of the Calzetti law ($p=0.7$ see also \citealt{cha00}). 

We also find a clear trend that the bump amplitude reduces as $A_{FUV}$
increases (panel [b]); from a similar value to that in the extinction
law to a very small amplitude even less than that of the LMC 2 (D)
case. Loci of the bump amplitudes along $A_{FUV}$ are very similar,
except for the LMC 2 (D) case whose amplitude is very small even in its
extinction law. This trend is also consistent with the literature 
\citep[e.g.,][]{var99,fer99,wg00,pie04,pan05}. The reduction of the bump
amplitude is originated from the nature of the grey attenuation in an
opaque disc as discussed in \S3.1. \cite{bur05b} did not find the
reduction of the bump amplitude along the FUV attenuation from their
analysis of the galaxies observed with the {\it GALEX}. This may be
because their sample consists of two different populations.
Indeed, the bump amplitudes estimated in their analysis show a bimodal
distribution.

\begin{figure}
 \centering
 \includegraphics[width=8cm]{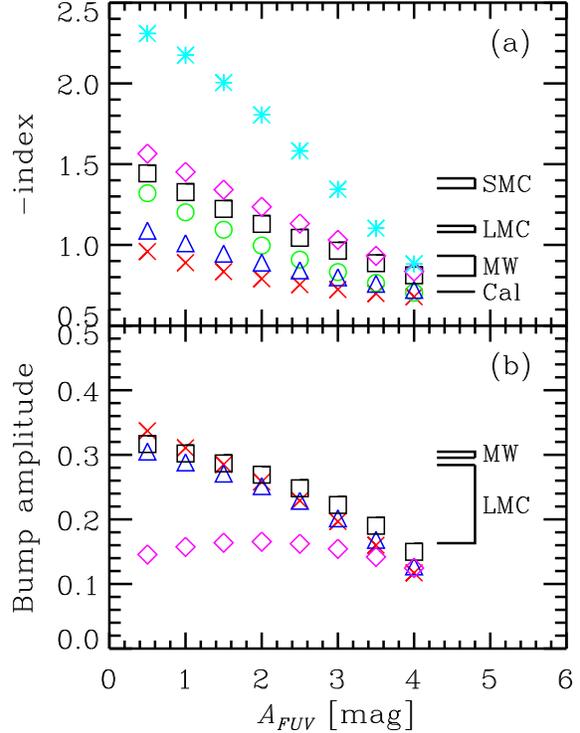}
 \caption{Properties of mean attenuation laws: (a) global slope of the 
 attenuation law and (b) amplitude of the bump. The vertical axes are
 defined in equation (6): index = $-p$ and bump amplitude = $B$. The
 crosses and circles are MW (WG) and SMC (WG), respectively. The
 triangles, squares, diamonds, and asterisks are MW (D), LMC av (D), LMC
 2 (D), and SMC (D), respectively. For a reference, the parameters of
 the extinction laws and the Calzetti law are marked in the right-hand
 side in each panel. Two marks for each extinction law mean two MW laws
 of WG and Drain dusts, two LMC laws of the Draine dust, and two SMC
 laws of WG and Draine dusts.}
\end{figure}

From Fig.~11, we understand why some dust models can reproduce the
{\it GALEX} data in Fig.~8 and other can not. The slope of the
attenuation law of the SMC (WG) is very good for the data. Suppose this
as the reference case. Although the LMC av (D) and the LMC 2 (D) give a
steeper slope than that of the SMC (WG), the bump which shifts the 
{\it GALEX} colour somewhat bluewards compensates the steepness. 
The slopes of the two MW dust cases are shallower than the reference SMC
(WG) case, and moreover, there is the bump. For the SMC (D) which does
not have the bump, the slope of the attenuation law is too steep.

\subsection{Relations between UV attenuation and other quantities}

We have derived the mean attenuation laws as a function of the FUV
attenuation, $A_{FUV}$. If we have $A_{FUV}$, thus, we can obtain an 
attenuation law from the UV to the NIR. Now, we need to know relations
between $A_{FUV}$ and other observable or theoretical quantities in
order to estimate $A_{FUV}$.

\subsubsection{IR-to-UV flux ratio}

\begin{figure}
 \centering
 \includegraphics[width=8cm]{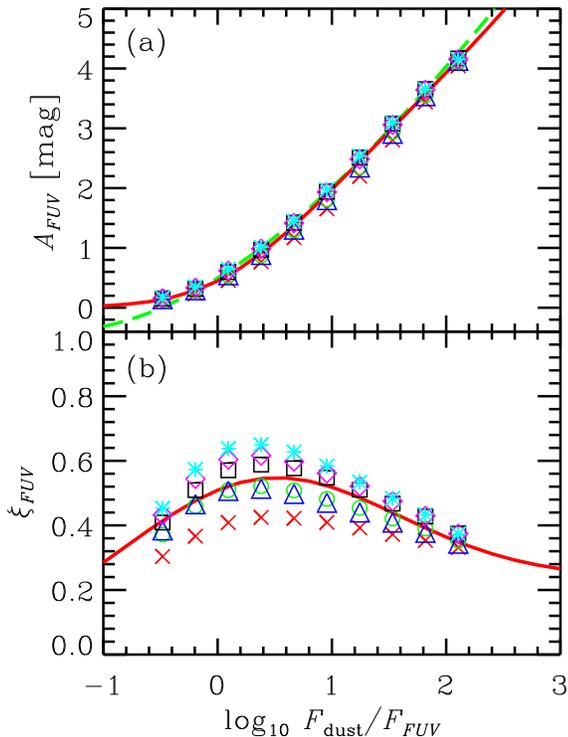}
 \caption{Mean relations between the IR-to-UV flux ratio and (a) the FUV
 attenuation and (b) the fraction of the FUV absorbed energy in the IR
 flux. The symbols, which almost overlap in the panel (a), indicate the
 mean relations for each dust model; crosses, circles, triangles,
 asterisks, squares, and diamonds are MW (WG), SMC (WG), MW (D), SMC
 (D), LMC av (D), and LMC 2 (D), respectively. The solid lines are the
 best fit curves expressed in equations (7) and (9). The dashed line in
 the panel (a) is the calibration proposed by Buat et al.~(2005).}
\end{figure}

The IR-to-UV flux ratio is a very good measure of 
$A_{FUV}$ \citep{bua96,gor00}. This is because the relation is based on
the energy conservation; the IR flux is the flux absorbed by dust grains.
Thus, the relation is very robust against differences of the
configuration of stars and dust and the dust model. Indeed, Fig.~12 (a)
shows the robustness of the relation against differences among dust
models. The standard deviations are less than 0.2 mag.
Since the FUV attenuation is defined as 
$A_{FUV}=2.5\log(F_{FUV}^{\rm int}/F_{FUV})$ with $F_{FUV}^{\rm int}$ 
and $F_{FUV}$ being the intrinsic and observed FUV fluxes, respectively, 
we can express the relation as follows \citep[e.g.,][]{meu99}:
\begin{equation}
 A_{FUV} = 2.5 \log 
  \left(1+\xi_{FUV}\frac{F_{\rm dust}}{F_{FUV}}\right)\,, 
\end{equation}
where
\begin{equation}
 \xi_{FUV} = \frac{F_{FUV}^{\rm int}-F_{FUV}}{F_{\rm dust}}\,,
\end{equation}
which is the energy fraction of the absorbed FUV flux in the IR flux.
From Fig.~12 (b), we find that $\xi_{FUV}$ is 0.3--0.6 although the
value depends on the dust model and the dispersion is somewhat large.
In an extremely dust poor case ($F_{\rm dust}/F_{FUV}\to0$), 
$F_{FUV}^{\rm int}=F_{FUV}$ but still $F_{\rm dust}>0$ because dust 
grains can be exposed by ionizing photons.
Thus, $\xi_{FUV}\to0$ when $F_{\rm dust}/F_{FUV}\to0$. 
In the opposite limit, all stellar radiation is absorbed by dust and 
re-emitted in the IR ($F_{FUV}\to0$ and $F_{\rm dust}\to F_{\rm total}$, 
where $F_{\rm total}$ is the total flux); 
$\xi_{FUV}\to\xi_{FUV}^\infty\equiv F_{FUV}^{\rm int}/F_{\rm total}$. 
If we assume the SED of a smooth exponential SFH with an e-folding time
of 5 Gyr and a galactic age of 10 Gyr, we have $\xi_{FUV}^\infty=0.245$.
With these limits, we have found a fitting function as 
\begin{equation}
 \xi_{FUV} = \frac{\xi_{FUV}^\infty e^x}{1+e^x}
  +0.400 \exp(-(0.591x-0.185)^2)\,,
\end{equation}
where $x=\log(F_{\rm dust}/F_{FUV})$. This is shown as the solid curve
in Fig.~12 (b). The relation obtained from equations (7) and (9) is
also shown as solid curve in Fig.~12 (a). The difference of the
calibration obtained here from that of \cite{bua05} (dashed curve in
Fig.~12 [a]) is very small, except for the region of a small 
$F_{\rm dust}/F_{FUV}$. Importantly, we can obtain a UV-to-NIR
attenuation law from the observed IR-to-UV flux ratio with the
calibration obtained here.

\subsubsection{UV colour}

\begin{figure}
 \centering
 \includegraphics[width=8cm]{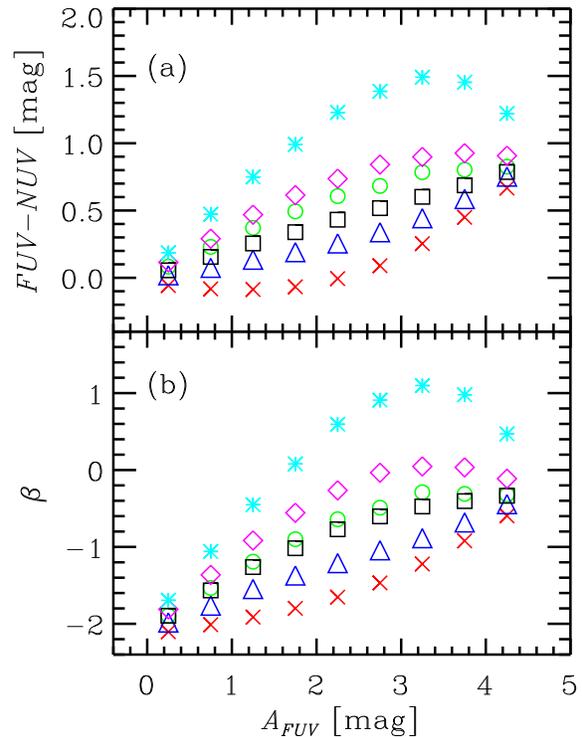}
 \caption{Mean relation between the FUV attenuation and the UV colour:
 (a) {\it GALEX} colour and (b) UV spectral slope $\beta$. The symbols
 are the same as Fig.~12.}
\end{figure}

As found by \cite{meu99}, the UV bright starburst galaxies show a tight
correlation between the UV spectral slope (i.e. UV colour) and the 
IR-to-UV flux ratio (i.e. $A_{FUV}$). However, normal galaxies do not
follow the relation \citep{bel02,kon04,bua05}. Figs.~5 and 6 show that the
relation between $A_{FUV}$ and the {\it GALEX} colour excess strongly
depends on the dust model. Thus, we expect that the UV colour and slope
also depend on the dust model significantly. We estimated
the UV slope $\beta$, where $f_\lambda \propto \lambda^\beta$, 
from the expected flux densities at 10 wavelength points between 0.12
and 0.26 \micron\ listed in table 2 of \cite{cal94}. As shown in
Fig.~13, we find that relations between $A_{FUV}$ and UV colour (or
slope) show a large dispersion depending on the dust model. For example,
$A_{FUV}$ distributes from 0.5 to 4.0 mag for $FUV-NUV=0.5$ mag or
$\beta=-1$. Although we may expect a smaller dispersion of $A_{FUV}$ for
a fixed UV colour (or slope) in each dust model, the dispersion is still
$\sim1$ mag.

\subsubsection{Dust column density}

\begin{figure}
 \centering
 \includegraphics[width=8cm]{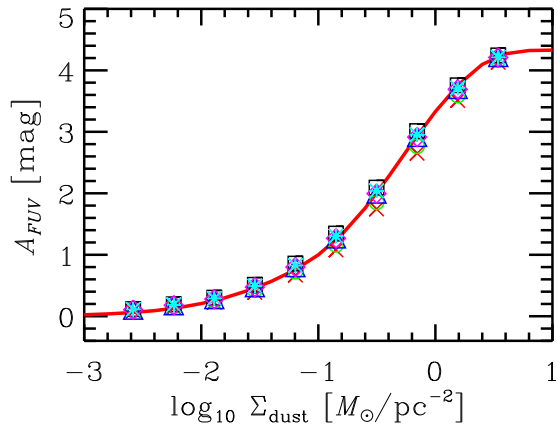}
 \caption{Mean relation between the FUV attenuation and the dust column
 density. The symbols are the same as Fig.~12. The solid line is the
 best fit curve described in equation (10).}
\end{figure}

To predict the observable fluxes of galaxies by theoretical models, 
we need a relation between $A_{FUV}$ and a theoretical quantity. 
Since the total dust column density, $\Sigma_{\rm dust}$, of the disc
can be  calculated by a galactic chemical and dust amount evolution
model \citep[e.g.,][]{ino03}, a relation between $A_{FUV}$ and
$\Sigma_{\rm dust}$ would be useful. 
Fig.~14 shows the $A_{FUV}$--$\Sigma_{\rm dust}$ relation which
is very robust against differences among dust models like the
relation between $A_{FUV}$ and the IR-to-UV flux ratio. 
A typical standard deviation is as small as 0.2 mag.
As $\Sigma_{\rm dust}$ increases, first $A_{FUV}$ increases linearly,
and then, becomes saturated at a certain value determined by the amount
of the source outside of the disc, i.e. layering parameter (see \S3.1). 
Taking into account such a behavior, we fit the data by the following
function: 
\begin{equation}
 A_{FUV} = \sum_{i=1}^3 a_i \{1-\exp(-b_i \Sigma_{\rm dust})\}\,.
\end{equation}
The resulting parameters are tabulated in Table~5, and the best fit
curve is shown as the solid line in Fig.~14.

\begin{table}
 \caption[]{Parameters for equation (10).}
 \setlength{\tabcolsep}{3pt}
 \footnotesize
 \begin{minipage}{\linewidth}
  \begin{tabular}{lcc}
   \hline
   $i$ & $a_i$ & $b_i$\\
   \hline
   1 & 0.299 & 55.3\\
   2 & 2.44 & 0.936\\
   3 & 1.59 & 3.62\\
   \hline
  \end{tabular}
 \end{minipage}
\end{table}%

\section{Conclusion}

We discussed dust properties in the ISM of nearby normal galaxies
observed through the {\it GALEX} FUV and NUV filters. To extract the
dust properties from the {\it GALEX} data, we built a huge set of
UV-to-NIR transmission rate curves of disc galaxies. The set covers a
very wide range of the physical parameters of these galaxies.
First, we examined various effects on the observed UV colour, such as 
the dust models, the age-selective attenuation including the clumpiness
of stars, the clumpiness of the dust distribution, and the SFH.
Next, we compared the distribution of the observational data of nearby
normal galaxies on the IR-to-UV flux ratio and the {\it GALEX} colour
diagram with that expected from our transmission rate curves. Then, 
we derived UV-to-NIR mean attenuation laws as a function of the FUV
attenuation from our transmission rate curves. We also derived 
relations between the FUV attenuation and the IR-to-UV flux ratio, the
UV colour, and the dust column density of the disc in order to use the
mean attenuation laws practically. 

From these analyses, we confirmed the following conclusions seen in the
literature: 
the 2175 \AA\ absorption bump, which is located in the {\it GALEX} NUV
filter, makes UV colours much bluer \citep[e.g.,][]{gor97,wg00}, 
the age-selective dust attenuation (i.e. younger stars are more
attenuated) reduces the bump strength in the {\it attenuation law} 
\citep{gra00,pan05}, the shape of the attenuation law changes from steep
to shallow as the FUV attenuation increases \citep[e.g.,][]{wg00}, and
the IR-to-UV flux ratio is a very good indicator of the FUV attenuation,
whereas the UV colour is not good one \citep[e.g.,][]{bua96,gor00}.

Our new findings are 
\begin{enumerate}
 \item[(1)] a rapid decline of the albedo toward a short
	    wavelength (except for the bump range) makes UV colours
	    significantly redder (\S3.1, 3.2), 

 \item[(2)] a smooth exponential SFH with an e-folding time-scale larger 
	    than 3 Gyr does not affect the UV colour excess
	    significantly (\S3.2), 

 \item[(3)] the FUV attenuation is a very good measure of the
	    attenuation at an arbitrary wavelength from the UV to the
	    NIR in our models (\S4.1), 

 \item[(4)] an age-dependent stellar feature like Mg I $\lambda2752$ can
	    appear in the attenuation law through the age-selective
	    attenuation (\S4.1), and 

 \item[(5)] the dust column density shows a very tight relation with
	    the FUV attenuation (\S4.3).
\end{enumerate}

From a comparison between our model and the {\it GALEX} observations in
\S3.3, we found that Witt \& Gordon's SMC type dust shows a very good
agreement with the {\it GALEX} data, whereas Witt \& Gordon's MW type
dust is not consistent with the data, and that Draine's LMC average and
LMC 2 type dusts are also well compatible with the {\it GALEX} data,
whereas Draine's SMC type dust is not consistent with the data. 
Draine's MW type dust gives slightly bluer colour than the data although
this discrepancy could be resolved easily, for example, with another
proper IMF as suggested by \cite{pan05}. Therefore, the main conclusion
of this paper is that 
\begin{enumerate}
 \item[(6)] in the ISM of the nearby normal galaxies, there is either 
	    dust with a bump and a smaller albedo for a shorter
	    wavelength (except for the bump range), or dust without any
	    bump but with an almost constant albedo.
\end{enumerate}
If we regard very small carbonaceous grains responsible for the UIR
emission band as the bump carrier \citep[e.g.,][]{whi03}, the former
dust is more suitable for nearby normal galaxies, because the UIR
emission is quite ubiquitous in these galaxies \citep{gen00}.

Explaining the behavior of the galaxies with a large IR-to-UV flux ratio
(typically $\ga100$) on the IR-to-UV flux ratio and the {\it GALEX}
colour diagram remains a future work. Our model predicts the convergence
to a certain {\it GALEX} colour independent of the dust model for
galaxies with such a large IR-to-UV flux ratio, whereas the nature shows
a very large dispersion of the colour. This may suggest a recent
episodic star formation, which decouples the stellar populations observed
in the UV and in the IR, in these galaxies.

\section*{acknowledgments}

We appreciate the simple stellar population spectra in electronic form
provided by Alessandro Bressan and useful comments from the anonymous
referee. 
AKI thanks Jean-Michel Deharveng for stimulating discussions and kind
encouragement, and Takako T. Ishii for providing her IDL functions which
are very useful for making figures. 
AKI also thanks all the members in the Laboratoire d'Astrophysique de
Marseille for their very kind hospitality during his stay there.
AKI and TTT were supported by the Japan Society for the Promotion of
Science (JSPS) Postdoctoral Fellowships for Research Abroad (AKI: Apr.\
2004--Mar.\ 2005; TTT: Apr.\ 2004--Dec.\ 2005).

\appendix

\section{Clumpy stellar emissivity}

Here we describe how to manage clumpiness of the stellar emissivity.
We do not consider any systematic distribution of clumps, i.e. clumps
distribute uniformly in the gas+dust disc. However, we consider that a
mean number density of embedded stars decreases along the vertical axis
of the disc. We update the treatment of the clumpy stellar emissivity of
the section 3.4.1 in \cite{ino05} as follows.  

First, we normalize the intrinsic emissivity, 
$\eta_*$, as unity. When clumps exist only within the gas+dust disc of
the half height $h_{\rm d}$ and the embedded stars exist only in clumps,
this normalization becomes 
\begin{equation}
\int_{-h_{\rm d}}^{h_{\rm d}}\eta_*(z)dz=1\,.
\end{equation}
If the number distribution of the embedded stars is an exponential
function along the vertical axis, $z$, with an e-folding scale of $h_*$,
we have $\eta_*(z)=\exp(-|z|/h_*)/2h_*(1-\exp(-h_{\rm d}/h_*))$ for 
$|z| \leq h_{\rm d}$, otherwise $\eta_*=0$.

A part of the radiation from the embedded stars is locally absorbed by
dust in the clump where the stars are embedded. If the embedded stars
distribute uniformly in the clump and the scattering by dust is
isotropic, the photon escape probability from the clump, 
$P_{\rm esc}(\tau_{\rm cl},\omega_{\rm cl})$, with the optical
depth radius $\tau_{\rm cl}$ and the effective albedo $\omega_{\rm cl}$, 
is given by equation (10) in \cite{ino05}. Therefore, the clumpy
stellar emissivity input into the transfer equation is 
\begin{equation}
  \eta_*^{\rm cl}(z) = \cases{
    \frac{P_{\rm esc}(\tau_{\rm cl},\omega_{\rm cl})
    \exp(-|z|/h_*)}{2h_*\{1-\exp(-h_{\rm d}/h_*)\}}
    & (for $|z| \leq h_{\rm d}$) \cr
    0 & (for $|z| > h_{\rm d}$) \cr}\,.
\end{equation}
In this case, we do not have any incident radiation at the top of the disc 
as the upper boundary condition of the calculation. In addition, we set a
mirror boundary condition at the equatorial plane of the disc.

\section{Estimation of the dust infrared luminosity}

Here, we explain the estimation procedure of the dust IR  
luminosity, in other words, the total absorbed luminosity.
The dust IR luminosity is expressed as \citep[e.g.,][]{ino00}
\begin{equation}
 L_{\rm dust} = L_{\rm Ly\alpha} + (1-f_{\rm LC}) L_{\rm LC} 
  + \int_{\lambda_{\rm L}}^\infty 
  L_\lambda (1-\langle T_\lambda \rangle) d\lambda\,,
\end{equation}
where $L_{\rm Ly\alpha}$ and $L_{\rm LC}$ are the Ly $\alpha$ emission
line and the Lyman continuum luminosities, $L_\lambda$ is the 
intrinsic luminosity density of all stellar populations, 
$f_{\rm LC}$ is the luminosity
fraction of the Lyman continuum absorbed by hydrogen atoms, and 
$\langle T_\lambda \rangle$ is the transmission rate averaged over the
angle between the disc normal and a ray. That is, 
\begin{equation}
 \langle T_\lambda \rangle = \int_0^1 T_\lambda (\mu) d\mu\,,
\end{equation}
$\mu$ is the cosine of the angle. Note that we should integrate
the {\it composite} transmission rate (with an angle dependence) over
the angle to obtain the total absorbed energy within the disc. We
assume the same SED as that used to obtain the luminosity weights in
equation (1) and Fig.~4 (solid lines). The first two terms in equation
(B1) can be reduced to $(1-\alpha f_{\rm LC}) L_{\rm LC}$ with $\alpha$
is a numerical factor of the order of unity \citep[e.g.,][]{ino00}. 
With the case B approximation and the assumed SED, we have 
$\alpha\simeq0.6$. Some estimates suggest $f_{\rm LC}\simeq0.5$
\citep{ihk01,ino01,ino02,hir03}. Thus, we adopt 
$1-\alpha f_{\rm LC}=0.7$. This choice does not affect the estimated
dust IR luminosity significantly because a typical luminosity fraction
of $L_{\rm LC}$ in the total luminosity is small enough (15\%).

\section{Parameters for mean attenuation laws}

We summarize the fitting parameters for equation (5) of each dust model
in Tables~C1 to C6.

\clearpage

\begin{table}
 \caption[]{Parameters for MW (WG)}
 \setlength{\tabcolsep}{3pt}
 \footnotesize
 \begin{minipage}{\linewidth}
  \begin{tabular}{lccc}
   \hline
   $\lambda$ (\micron) & $\alpha$ & $\beta$ & $\gamma$\\
   \hline
   0.100 &  1.443E+00 & -4.356E-01 &  3.637E-02\\
   0.114 &  7.122E-01 & -1.762E-01 &  1.066E-02\\
   0.129 &  2.658E-01 & -5.189E-02 &  1.559E-03\\
   0.143 &  5.388E-02 & -9.868E-03 &  1.126E-03\\
   0.157 & -4.344E-02 &  8.463E-03 & -1.066E-04\\
   0.171 & -6.372E-02 &  1.537E-02 & -2.479E-03\\
   0.186 & -6.457E-03 &  9.752E-03 & -5.950E-03\\
   0.200 &  1.872E-01 & -3.753E-02 & -7.192E-03\\
   0.214 &  4.086E-01 & -1.208E-01 & -1.885E-03\\
   0.229 &  2.693E-01 & -8.075E-02 & -5.494E-03\\
   0.243 &  4.049E-02 & -1.150E-02 & -9.710E-03\\
   0.257 & -1.145E-01 &  2.076E-02 & -1.076E-02\\
   0.271 & -2.049E-01 &  2.850E-02 & -1.158E-02\\
   0.286 & -2.642E-01 &  3.982E-02 & -1.117E-02\\
   0.300 & -3.191E-01 &  2.379E-02 & -1.160E-02\\
   0.378 & -4.347E-01 &  6.473E-03 & -8.980E-03\\
   0.475 & -5.609E-01 &  7.314E-03 & -7.126E-03\\
   0.599 & -6.473E-01 &  1.715E-02 & -6.257E-03\\
   0.754 & -7.206E-01 &  2.182E-02 & -4.276E-03\\
   0.949 & -7.969E-01 &  1.269E-02 &  9.968E-04\\
   1.194 & -8.273E-01 & -2.497E-03 &  5.293E-03\\
   1.504 & -8.508E-01 & -1.808E-02 &  8.990E-03\\
   1.893 & -8.700E-01 & -3.024E-02 &  1.135E-02\\
   2.383 & -8.848E-01 & -3.682E-02 &  1.219E-02\\
   3.000 & -8.984E-01 & -3.935E-02 &  1.202E-02\\
   \hline
  \end{tabular}
 \end{minipage}
\end{table}%

\begin{table}
 \caption[]{Parameters for SMC (WG)}
 \setlength{\tabcolsep}{3pt}
 \footnotesize
 \begin{minipage}{\linewidth}
  \begin{tabular}{lccc}
   \hline
   $\lambda$ (\micron) & $\alpha$ & $\beta$ & $\gamma$\\
   \hline
   0.100 &  1.207E+00 & -3.332E-01 &  2.472E-02\\
   0.114 &  6.343E-01 & -1.435E-01 &  7.019E-03\\
   0.129 &  2.923E-01 & -5.089E-02 & -7.859E-05\\
   0.143 &  9.794E-02 & -1.118E-02 & -7.718E-04\\
   0.157 & -4.001E-02 &  1.132E-02 & -8.986E-04\\
   0.171 & -1.454E-01 &  2.428E-02 & -4.576E-04\\
   0.186 & -2.226E-01 &  3.206E-02 & -2.444E-04\\
   0.200 & -2.871E-01 &  3.780E-02 & -3.013E-04\\
   0.214 & -3.344E-01 &  4.106E-02 & -7.320E-04\\
   0.229 & -3.786E-01 &  4.194E-02 & -3.141E-04\\
   0.243 & -4.069E-01 &  3.710E-02 &  1.830E-03\\
   0.257 & -4.429E-01 &  3.767E-02 &  1.865E-03\\
   0.271 & -4.820E-01 &  4.311E-02 & -8.637E-04\\
   0.286 & -5.139E-01 &  3.827E-02 &  1.631E-03\\
   0.300 & -5.605E-01 &  4.701E-02 & -4.466E-03\\
   0.378 & -6.488E-01 &  3.886E-02 & -5.286E-03\\
   0.475 & -7.278E-01 &  2.905E-02 & -3.543E-03\\
   0.599 & -7.841E-01 &  1.859E-02 &  1.004E-04\\
   0.754 & -8.222E-01 &  4.153E-03 &  4.159E-03\\
   0.949 & -8.521E-01 & -1.531E-02 &  8.775E-03\\
   1.194 & -8.728E-01 & -3.147E-02 &  1.182E-02\\
   1.504 & -8.927E-01 & -4.005E-02 &  1.268E-02\\
   1.893 & -9.059E-01 & -4.106E-02 &  1.204E-02\\
   2.383 & -9.178E-01 & -3.951E-02 &  1.089E-02\\
   3.000 & -9.379E-01 & -3.338E-02 &  8.090E-03\\
   \hline
  \end{tabular}
 \end{minipage}
\end{table}%

\begin{table}
 \caption[]{Parameters for MW (D)}
 \setlength{\tabcolsep}{3pt}
 \footnotesize
 \begin{minipage}{\linewidth}
  \begin{tabular}{lccc}
   \hline
   $\lambda$ (\micron) & $\alpha$ & $\beta$ & $\gamma$\\
   \hline
   0.100 &  8.548E-01 & -2.088E-01 &  1.413E-02\\
   0.114 &  4.825E-01 & -1.044E-01 &  5.833E-03\\
   0.129 &  2.151E-01 & -4.025E-02 &  1.456E-03\\
   0.143 &  8.084E-02 & -1.464E-02 &  9.074E-04\\
   0.157 & -3.413E-02 &  7.360E-03 & -2.950E-04\\
   0.171 & -1.147E-01 &  2.546E-02 & -2.297E-03\\
   0.186 & -1.448E-01 &  4.049E-02 & -6.133E-03\\
   0.200 & -7.306E-02 &  3.724E-02 & -1.109E-02\\
   0.214 &  5.246E-02 &  5.592E-03 & -1.254E-02\\
   0.229 & -6.209E-02 &  3.145E-02 & -1.423E-02\\
   0.243 & -2.465E-01 &  6.706E-02 & -1.321E-02\\
   0.257 & -3.671E-01 &  7.848E-02 & -1.139E-02\\
   0.271 & -4.389E-01 &  8.058E-02 & -1.184E-02\\
   0.286 & -4.776E-01 &  8.018E-02 & -9.885E-03\\
   0.300 & -5.207E-01 &  7.287E-02 & -1.274E-02\\
   0.378 & -6.643E-01 &  5.906E-02 & -9.598E-03\\
   0.475 & -7.638E-01 &  4.535E-02 & -6.066E-03\\
   0.599 & -8.198E-01 &  3.164E-02 & -1.767E-03\\
   0.754 & -8.487E-01 &  1.481E-02 &  2.467E-03\\
   0.949 & -8.626E-01 & -2.464E-03 &  6.332E-03\\
   1.194 & -8.707E-01 & -1.802E-02 &  9.473E-03\\
   1.504 & -8.769E-01 & -3.025E-02 &  1.170E-02\\
   1.893 & -8.835E-01 & -3.866E-02 &  1.296E-02\\
   2.383 & -8.918E-01 & -4.312E-02 &  1.328E-02\\
   3.000 & -9.019E-01 & -4.409E-02 &  1.281E-02\\
   \hline
  \end{tabular}
 \end{minipage}
\end{table}%

\begin{table}
 \caption[]{Parameters for LMC av (D)}
 \setlength{\tabcolsep}{3pt}
 \footnotesize
 \begin{minipage}{\linewidth}
  \begin{tabular}{lccc}
   \hline
   $\lambda$ (\micron) & $\alpha$ & $\beta$ & $\gamma$\\
   \hline
   0.100 &  1.140E+00 & -3.017E-01 &  2.097E-02\\
   0.114 &  6.430E-01 & -1.445E-01 &  6.918E-03\\
   0.129 &  3.107E-01 & -5.632E-02 &  3.382E-04\\
   0.143 &  1.327E-01 & -1.758E-02 & -9.865E-04\\
   0.157 & -5.417E-02 &  1.508E-02 & -1.054E-03\\
   0.171 & -1.805E-01 &  3.349E-02 & -8.935E-04\\
   0.186 & -2.348E-01 &  4.734E-02 & -3.223E-03\\
   0.200 & -1.839E-01 &  5.265E-02 & -8.902E-03\\
   0.214 & -7.520E-02 &  3.457E-02 & -1.243E-02\\
   0.229 & -1.851E-01 &  5.198E-02 & -1.240E-02\\
   0.243 & -3.571E-01 &  6.668E-02 & -7.398E-03\\
   0.257 & -4.713E-01 &  6.447E-02 & -2.996E-03\\
   0.271 & -5.439E-01 &  6.472E-02 & -3.028E-03\\
   0.286 & -5.781E-01 &  5.665E-02 &  2.720E-04\\
   0.300 & -6.281E-01 &  6.490E-02 & -5.516E-03\\
   0.378 & -7.637E-01 &  4.383E-02 & -1.448E-03\\
   0.475 & -8.453E-01 &  1.703E-02 &  3.602E-03\\
   0.599 & -8.781E-01 & -1.221E-02 &  9.342E-03\\
   0.754 & -8.903E-01 & -3.086E-02 &  1.237E-02\\
   0.949 & -8.952E-01 & -4.055E-02 &  1.361E-02\\
   1.194 & -8.994E-01 & -4.538E-02 &  1.391E-02\\
   1.504 & -9.034E-01 & -4.714E-02 &  1.372E-02\\
   1.893 & -9.065E-01 & -4.725E-02 &  1.333E-02\\
   2.383 & -9.094E-01 & -4.650E-02 &  1.284E-02\\
   3.000 & -9.141E-01 & -4.493E-02 &  1.214E-02\\
   \hline
  \end{tabular}
 \end{minipage}
\end{table}%

\begin{table}
 \caption[]{Parameters for LMC 2 (D)}
 \setlength{\tabcolsep}{3pt}
 \footnotesize
 \begin{minipage}{\linewidth}
  \begin{tabular}{lccc}
   \hline
   $\lambda$ (\micron) & $\alpha$ & $\beta$ & $\gamma$\\
   \hline
   0.100 &  9.849E-01 & -2.498E-01 &  1.699E-02\\
   0.114 &  5.809E-01 & -1.278E-01 &  6.279E-03\\
   0.129 &  3.089E-01 & -5.578E-02 &  3.272E-04\\
   0.143 &  1.549E-01 & -2.023E-02 & -1.482E-03\\
   0.157 & -4.878E-02 &  1.837E-02 & -2.057E-03\\
   0.171 & -2.238E-01 &  4.130E-02 & -5.816E-04\\
   0.186 & -3.441E-01 &  5.403E-02 &  4.379E-04\\
   0.200 & -3.839E-01 &  6.664E-02 & -2.391E-03\\
   0.214 & -3.658E-01 &  7.510E-02 & -7.153E-03\\
   0.229 & -4.411E-01 &  7.418E-02 & -4.840E-03\\
   0.243 & -5.344E-01 &  5.563E-02 &  3.301E-03\\
   0.257 & -5.998E-01 &  4.137E-02 &  7.775E-03\\
   0.271 & -6.490E-01 &  4.151E-02 &  6.645E-03\\
   0.286 & -6.673E-01 &  2.952E-02 &  9.911E-03\\
   0.300 & -7.130E-01 &  4.771E-02 &  1.951E-03\\
   0.378 & -8.037E-01 &  2.841E-02 &  3.376E-03\\
   0.475 & -8.581E-01 &  7.700E-03 &  5.843E-03\\
   0.599 & -8.795E-01 & -1.402E-02 &  9.644E-03\\
   0.754 & -8.883E-01 & -2.872E-02 &  1.194E-02\\
   0.949 & -8.928E-01 & -3.802E-02 &  1.319E-02\\
   1.194 & -8.973E-01 & -4.311E-02 &  1.359E-02\\
   1.504 & -9.019E-01 & -4.537E-02 &  1.349E-02\\
   1.893 & -9.041E-01 & -4.566E-02 &  1.320E-02\\
   2.383 & -9.065E-01 & -4.516E-02 &  1.280E-02\\
   3.000 & -9.120E-01 & -4.371E-02 &  1.207E-02\\
   \hline
  \end{tabular}
 \end{minipage}
\end{table}%

\begin{table}
 \caption[]{Parameters for SMC (D)}
 \setlength{\tabcolsep}{3pt}
 \footnotesize
 \begin{minipage}{\linewidth}
  \begin{tabular}{lccc}
   \hline
   $\lambda$ (\micron) & $\alpha$ & $\beta$ & $\gamma$\\
   \hline
   0.100 &  1.043E+00 & -2.622E-01 &  1.688E-02\\
   0.114 &  6.491E-01 & -1.392E-01 &  5.393E-03\\
   0.129 &  3.831E-01 & -6.501E-02 & -1.364E-03\\
   0.143 &  2.148E-01 & -2.229E-02 & -4.045E-03\\
   0.157 & -5.989E-02 &  3.263E-02 & -4.722E-03\\
   0.171 & -3.110E-01 &  5.495E-02 &  5.624E-04\\
   0.186 & -4.910E-01 &  4.419E-02 &  9.322E-03\\
   0.200 & -5.982E-01 &  2.121E-02 &  1.698E-02\\
   0.214 & -6.509E-01 &  5.248E-03 &  2.060E-02\\
   0.229 & -6.788E-01 & -9.267E-03 &  2.328E-02\\
   0.243 & -6.912E-01 & -3.038E-02 &  2.766E-02\\
   0.257 & -7.093E-01 & -3.737E-02 &  2.829E-02\\
   0.271 & -7.358E-01 & -2.737E-02 &  2.429E-02\\
   0.286 & -7.384E-01 & -3.690E-02 &  2.616E-02\\
   0.300 & -7.807E-01 & -4.948E-03 &  1.573E-02\\
   0.378 & -8.308E-01 & -1.032E-03 &  1.058E-02\\
   0.475 & -8.621E-01 & -4.739E-03 &  8.648E-03\\
   0.599 & -8.740E-01 & -1.698E-02 &  1.029E-02\\
   0.754 & -8.810E-01 & -2.857E-02 &  1.204E-02\\
   0.949 & -8.860E-01 & -3.811E-02 &  1.341E-02\\
   1.194 & -8.912E-01 & -4.398E-02 &  1.398E-02\\
   1.504 & -8.970E-01 & -4.654E-02 &  1.390E-02\\
   1.893 & -9.034E-01 & -4.678E-02 &  1.339E-02\\
   2.383 & -9.100E-01 & -4.562E-02 &  1.263E-02\\
   3.000 & -9.166E-01 & -4.365E-02 &  1.175E-02\\
   \hline
  \end{tabular}
 \end{minipage}
\end{table}%

\label{lastpage}

\end{document}